\documentclass[sigconf]{acmart}
\AtBeginDocument{%
  \providecommand\BibTeX{{%
    \normalfont B\kern-0.5em{\scshape i\kern-0.25em b}\kern-0.8em\TeX}}}

\copyrightyear{2024}
\acmYear{2024}
\setcopyright{rightsretained}
\acmConference[CHI '24]{Proceedings of the CHI Conference on Human Factors in Computing Systems}{May 11--16, 2024}{Honolulu, HI, USA}
\acmBooktitle{Proceedings of the CHI Conference on Human Factors in Computing Systems (CHI '24), May 11--16, 2024, Honolulu, HI, USA}
\acmDOI{10.1145/3613904.3642761}
\acmISBN{979-8-4007-0330-0/24/05}

\acmConference[CHI'24]{Make sure to enter the correct
  conference title from your rights confirmation emai}{May 11--16,
  2024}{Honolulu, HI, USA}
\acmBooktitle{CHI'24: ACM CHI conference on Human Factors in Computing Systems, May 11--16,
  2024, Hawaii, USA} 
\acmPrice{15.00}
\acmISBN{979-8-4007-0330-0/24/05}

\usepackage{tikz}
\usepackage{soul}
\usepackage{url}
\usepackage[utf8]{inputenc}
\usepackage{graphicx}
\usepackage{amsmath}
\usepackage{amsthm}
\usepackage{booktabs}
\usepackage{algorithm}
\usepackage{algorithmic}
\usepackage{float}
\usepackage{array}

\graphicspath{{figures/}}
\usepackage{subfig}
\usepackage{multirow}
\usepackage{bm}
\usepackage{color}
\definecolor{Gray}{gray}{0.935}

\begin{document}


\newcommand{\ashish}[1]{\textcolor{red}{#1-Ashish}}
\newcommand{\inna}[1]{\textcolor{blue}{#1-Inna}}
\newcommand{\tim}[1]{\textcolor{green}{#1-Tim}}

\newcommand{\xhdr}[1]{\vspace{1.6mm}\noindent{{\bf #1.}}}

\definecolor{valbest}{HTML}{d9ead3}
\newcommand{\valbest}[1]{\colorbox{valbest}{#1}}
\definecolor{valgood}{HTML}{d0e0e3}
\newcommand{\valgood}[1]{\colorbox{valgood}{#1}}
\definecolor{valmid}{HTML}{fce5cd}
\newcommand{\valmid}[1]{\colorbox{valmid}{#1}}
\definecolor{valbad}{HTML}{ead1dc}
\newcommand{\valbad}[1]{\colorbox{valbad}{#1}}

\newcommand{\valingood}[1]{\begingroup\setlength{\fboxsep}{2pt}
\colorbox{valgood}{#1}
\endgroup
}

\newcommand{\valinbad}[1]{\begingroup\setlength{\fboxsep}{2pt}
\colorbox{valbad}{#1}
\endgroup
}

\newcommand{\Hone}{$\mathbf{H_1}$}
\newcommand{\Htwo}{$\mathbf{H_2}$}
\newcommand{\Hthree}{$\mathbf{H_3}$}
\newcommand{\Hfour}{$\mathbf{H_4}$}
\newcommand{\Hfive}{$\mathbf{H_5}$}

\newcolumntype{M}[1]{>{\centering\arraybackslash}m{#1}}

\newcommand{\rone}[1]{{#1}}
\newcommand{\rtwo}[1]{{#1}}
\newcommand{\rthree}[1]{{#1}}

\title[Facilitating Self-Guided Mental Health Interventions Through Human-Language Model Interaction]{Facilitating Self-Guided Mental Health Interventions Through \mbox{Human-Language Model Interaction:} A Case Study of Cognitive Restructuring}


\author{Ashish Sharma}
\email{ashshar@cs.washington.edu}
\affiliation{%
  \institution{University of Washington}
  \city{Seattle}
  \state{WA}
  \country{USA}
}

\author{Kevin Rushton}
\email{krushton@mhanational.org}
\affiliation{%
  \institution{Mental Health America}
  \city{Alexandria}
  \state{VA}
  \country{USA}
}

\author{Inna Wanyin Lin}
\email{ilin@cs.washington.edu}
\affiliation{%
  \institution{University of Washington}
  \city{Seattle}
  \state{WA}
  \country{USA}
}

\author{Theresa Nguyen}
\email{tnguyen@mhanational.org }
\affiliation{%
  \institution{Mental Health America}
  \city{Alexandria}
  \state{VA}
  \country{USA}
}

\author{Tim Althoff}
\email{althoff@cs.washington.edu}
\affiliation{%
  \institution{University of Washington}
  \city{Seattle}
  \state{WA}
  \country{USA}
}

\renewcommand{\shortauthors}{Sharma et al.}





\begin{abstract}
  Self-guided mental health interventions, such as ``do-it-yourself'' tools to learn and practice coping strategies, show great promise to improve access to mental health care. However, these interventions are often cognitively demanding and emotionally triggering, creating accessibility barriers that limit their wide-scale implementation and adoption. In this paper, we study how human-language model interaction can support self-guided mental health interventions. We take \textit{cognitive restructuring}, an evidence-based therapeutic technique to overcome negative thinking, as a case study. In an IRB-approved randomized field study on a large mental health website with 15,531 participants, we design and evaluate a system that uses language models to support people through various steps of cognitive restructuring. Our findings reveal that our system positively impacts emotional intensity for 67\% of participants and helps 65\% overcome negative thoughts. Although adolescents report relatively worse outcomes, we find that tailored interventions that simplify language model generations improve overall effectiveness and equity.
\end{abstract}

\maketitle

\section{Introduction}
As mental health conditions surge worldwide, healthcare systems are struggling to provide accessible mental health care for all~\cite{world2022mental,olfson2016building,sickel2014mental}. Self-guided mental health interventions, such as tools to journal and reflect on negative thoughts, offer great promise to expand modes of care and help people learn coping strategies \cite{schleider2020future,patel2020acceptability,schleider2022randomized,shkel2023understanding}. While not a replacement for formal psychotherapy, these interventions provide immediate on-demand access to resources that can help develop techniques for mental well-being, especially for those who lack access to a trained professional, are on waiting lists, or seek to supplement therapy with other forms of care \cite{adair2005continuity}.

However, developing interventions that individuals can effectively use without the assistance of a professional therapist is challenging \cite{garrido2019works}. Currently, most self-guided interventions simply transform
traditional manual therapeutic worksheets into digital online formats with limited instructions and support \cite{shkel2023understanding}. Using these worksheets without professional support often leads to cognitively demanding and emotionally triggering experiences, limiting engagement and usage \cite{garrido2019works,baumel2019objective,fleming2018beyond,torous2020dropout}. For example, a popular self-guided intervention involves independently practicing \textit{Cognitive Restructuring of Negative Thoughts}, an evidence-based, well-established process that helps people notice and change their negative thinking patterns \cite{beck1976cognitive,burns1980feeling}. However, the practice includes complex steps like \textit{identifying thinking traps} (faulty or distorted patterns of thinking like ``\textit{all-or-nothing thinking}''), which pose a significant challenge for many \rthree{\cite{beck1976cognitive,burns1980feeling}}. Most individuals lack the necessary knowledge or experience to successfully use such interventions independently without explicit training and support. Moreover, analyzing one's own thoughts, emotions, and behavioral patterns can be emotionally triggering, especially for those actively experiencing distress. Such accessibility barriers inhibit the widespread adoption of self-guided mental health interventions.

Language models may be able to assist individuals in engaging with self-guided mental health interventions to improve intervention accessibility and effectiveness.
Specifically, language models can help individuals learn and practice cognitively demanding tasks (e.g., through automatic suggestions on potential thinking traps in a thought). Moreover, the language model support could help users manage emotionally triggering thoughts and potentially enable a reduction in the emotional intensity of negative thoughts.

Previous work exploring self-guided mental health interventions based on human-language model interaction has predominantly been limited to small-scale \cite{ly2017fully} and wizard-of-oz-style research \cite{smith2021effective,morris2015efficacy,kornfield2023text,kumar2023exploring}. These studies were conducted in controlled lab settings and evaluated on online crowdworker platforms, such as MTurk, which may not accurately represent the people who actively seek mental health care or use such an intervention \cite{mohr2017accelerating}. Much less is known about intervention effectiveness in ecologically valid settings with individuals experiencing mental health challenges and seeking care. This limits our understanding of end-user preferences within these emerging forms of intervention \cite{mohr2017accelerating,blandford2018seven,borghouts2021barriers,poole2013hci}. Furthermore, language models can exhibit biases resulting in highly varied performance across people from diverse demographics and populations \cite{caliskan2017semantics,blodgett2020language,lin-etal-2022-gendered}. There is a need to investigate and improve the equity of language modeling-based interventions.

In this paper, we study supporting self-guided mental health interventions with human-language model interaction. Specifically, we take \textit{Cognitive Restructuring}, an evidence-based self-guided intervention, as a case study. Cognitive Restructuring helps individuals to recognize when they are getting stuck in distorted patterns of thinking (\textit{identifying thinking traps}) and to come up with new ways to think about their situation (\textit{writing reframed thoughts}). 

We design and evaluate a human-language model interaction based tool for cognitive restructuring. We conduct an ecologically valid and large-scale randomized field study on Mental Health America (\href{https://screening.mhanational.org/}{screening.mhanational.org}; MHA; a popular website that hosts mental health tools and resources) with over 15,000 participants (Section~\ref{sec:study-overview}). We examine the design of our tool, investigate its impact on people seeking mental health care, and evaluate and improve its equity across key subpopulations. Specifically, we address the following research questions:

\xhdr{RQ1 – Design} How can we design a self-guided cognitive restructuring intervention that is supported through human-language model interaction?

\xhdr{RQ2a – Overall Effectiveness} To what extent does human-language model interaction based cognitive restructuring help individuals in alleviating negative emotions and overcoming negative thoughts?

\xhdr{RQ2b – Impact of Design Hypotheses} What is the impact of individual design hypotheses on the intervention effectiveness? 

\xhdr{RQ3 – Equity} How equitable is the intervention and what strategies may improve equity?

To address these research questions, we first formulate the design hypotheses for this intervention through qualitative feedback from participants of early prototypes of the system and brainstorming with mental health professionals (Section~\ref{subsec:hypothesis}). \rthree{We hypothesize that assisting users in cognitively and emotionally challenging processes, contextualizing thought reframes by reflecting on situations and emotions, integrating psychoeducation, facilitating interactive refinement of reframes, and ensuring safety will result in a more effective human-language model interaction system for cognitive restructuring}. Taking these hypotheses into consideration, we design a new system that uses a language model to support people through various steps of cognitive restructuring, including the \textit{identification} of thinking traps in thoughts and the \textit{reframing} of negative thoughts. Our language model suggests possible thinking traps a given thought may have, as well as suggests possible ways of reframing negative thought (Section~\ref{subsec:system}).

After systematic \rthree{ethical} and safety considerations, active collaboration with mental health professionals, advocates, and clinical psychologists, and IRB review and approval, we deploy this system on MHA. We utilize a mixed-methods study design, combining quantitative and qualitative feedback, to assess the outcomes of this system on the platform visitors. We find that 67.64\% of participants experience a positive shift (i.e., reduction) in their emotion intensity and 65.65\% report helpfulness in overcoming their negative thoughts through the use of our system. Moreover, participants indicate that the system assists them in alleviating cognitive barriers by simplifying task complexity and emotional barriers by providing a less triggering experience (Section~\ref{sec:effectiveness-overall}). Also, enabling participants to iteratively improve reframes by seeking additional reframing suggestions leads to a 23.73\% greater reduction in the intensity of negative emotions. Moreover, participants who choose to make their reframes actionable report superior outcomes compared to those who make them empathic or personalized (Section~\ref{sec:impact-design-hypothesis}).

To address the needs of individuals from diverse demographics and subpopulations, it is critical to develop equitable solutions. Here, we evaluate the performance of our system across people of different demographics and subpopulations (Section~\ref{sec:equity}). The intervention is found to be more effective for individuals identifying as females, older adults, individuals with higher education levels, and those struggling with \rthree{issues, such as parenting and work}. However, it is found to be less effective for individuals identifying as males, adolescents, those with lower education levels, and those struggling with issues such as hopelessness or loneliness. We investigate the potential benefits of customizing the intervention for adolescents, who we found to have one of the largest disparities in our intervention outcomes. We find that making the language model-generated suggestions simpler and more casual leads to a 14.44\% increase in reframe helpfulness for adolescents in the age group of 13 to 14.  

We discuss the implications of our study for the use of human-language interaction in the development of self-guided mental health interventions, emphasizing the need for effective personalization, appropriate levels of interactivity with the language model, equity across subpopulations, and its safety (Section~\ref{sec:discussion}). 

\rtwo{To summarize, our contributions are as follows:}
\rtwo{\begin{enumerate}
    \item Design of a novel health and wellness technology for human-language model interaction based self-guided mental health intervention.
    \begin{itemize}
        \item We \textbf{design and evaluate a new system} that uses language models to help people through cognitive restructuring of negative thoughts, a core, well-established intervention in Cognitive Behavioral Therapy (Section~\ref{sec:design}).
        \item We \textbf{formulate design hypotheses/decisions} that researchers developing self-guided interventions must consider to increase efficacy and ensure safety (Section~\ref{sec:design}).
        \item We \textbf{conduct randomized controlled trials} to assess the impact of different design hypotheses/decisions (Sections~\ref{sec:effectiveness-overall} and \ref{sec:impact-design-hypothesis}), how equitable the technology is, and what strategies may improve its equity (Section~\ref{sec:equity}).
    \end{itemize}
    \item Large-scale, randomized, empirical studies in an ecologically informed setting to understand how people with lived experience of mental health interact with this technology and key takeaways for researchers developing self-guided interventions that leverage human-language model interaction. 
    \begin{itemize}
        \item \textbf{There are opportunities to assist users in cognitively challenging and emotionally triggering psychological processes:} Effective self-guided mental health interventions could rapidly increase access to care. However, there are many steps in such interventions that pose cognitive and emotional challenges, leading to well-documented, high rates of participant dropout. Our work shows how the use of human-language model interaction can help address these challenges and help people use these interventions more effectively (Section~\ref{sec:effectiveness-overall}). 
        \item \textbf{Certain types of suggestions are most effective: }Our study suggests that users are more likely to seek more actionable suggestions and are more likely to find actionable suggestions more helpful (Section~\ref{sec:impact-design-hypothesis}).
        \item \textbf{Human-language model interaction systems are not necessarily equitable, but it is possible to adapt them to individual subgroups:} Our work shows that human-language model interaction systems are likely to have heterogeneous effects across key subpopulations that require further adapting the intervention (Section~\ref{sec:equity}).
    \end{itemize}
\end{enumerate}
}
\section{Related Work}
\label{sec:related-work}

Our work builds upon previous research on digital mental health interventions (Section~\ref{subsec:related-self-guided}), AI for mental health (Section~\ref{subsec:related-ai-mental-health}), and the design of human-language model interaction systems (Section~\ref{subsec:related-human-language-model-interaction}).

\subsection{Digital Mental Health Interventions}
\label{subsec:related-self-guided}
The critical gap between the overwhelming need for and limited access to mental health care has prompted clinicians, technologists, and advocates to develop digital interventions that provide accessible care for all. Several efforts have concentrated on facilitating digital, text-based supportive conversations, through peer-to-peer support networks, such as TalkLife (\href{https://www.talklife.com/}{talklife.com}) and Supportiv (\href{https://www.supportiv.com/}{supportiv.com}), as well as through on-demand talk therapy platforms like Talkspace (\href{https://www.talkspace.com/}{talkspace.com}), BetterHelp (\href{betterhelp.com}{betterhelp.com}), and SanVello (\href{https://www.sanvello.com/}{sanvello.com}). Researchers have conducted randomized controlled trials to study the efficacy of these interventions compared to traditional methods of care, such as in-person counseling and worksheet-based skill practices \cite{hull2017study,song2023comparing,moberg2019guided}.

Another key focus in this pursuit has been the development of self-guided mental health interventions \cite{schleider2020future,patel2020acceptability,schleider2022randomized,shkel2023understanding}. These interventions are designed in various forms, such as ``\textit{Do-It-Yourself}'' apps to improve mental health ``in-the-moment'' of crisis, self-help tools for learning and practicing therapeutic skills, and more. Popular examples include self-guided meditation such as Headspace (\href{https://www.headspace.com}{headspace.com}) or Calm (\href{https://www.calm.com/}{calm.com}). Researchers have also explored the design of apps to track mood changes \cite{schueller2021understanding}, emotion regulation \cite{smith2022digital} and to combat loneliness \cite{boucher2021impact}. Kruzan et al. \cite{kruzan2022wanted} studied the process of online self-screening of mental illnesses and its role in help-seeking. Howe et al. \cite{howe2022design} designed and evaluated a workplace stress-reduction intervention system and found that high-effort interventions reduced the most stress. Also, Cognitive Behavioral Therapy (CBT) is an evidence-based, well-established psychological treatment \cite{beck1976cognitive}. Researchers have designed digital self-guided interventions that streamline elements of CBT like cognitive restructuring of negative thoughts by transforming traditional manual worksheets into digital online formats \cite{rennick2016health,shkel2023understanding}. Other work has focused on digital mental health interventions based on Dialectical Behavioral Therapy (DBT) \cite{schroeder2018pocket,schroeder2020data}.

Our work extends this literature by investigating how to design and evaluate self-guided digital mental health interventions that leverage human-language model interaction. We take CBT-based cognitive restructuring as a case study and design and evaluate a system that assists individuals in restructuring of negative thoughts through language models in a large-scale randomized field study.

\subsection{AI for Mental Health}
\label{subsec:related-ai-mental-health}

Our work is related to the growing field of research in Artificial Intelligence (AI) and Natural Language Processing (NLP) for mental health and wellbeing.

Prior work has proposed \rthree{machine learning-based} methods to measure key mental health constructs including 
adaptability and efficacy of counselors \cite{perez2019makes}, personalized vs templated counseling language~\cite{althoff2016large}, 
psychological perspective change~\cite{althoff2016large,wadden2021moderation}, 
staying on topic~\cite{wadden2021moderation},  
therapeutic actions \cite{lee2019identifying}, empathy \cite{sharma2020computational}, moments of change \cite{pruksachatkun2019moments}, counseling strategies \cite{perez2022pair,shah2022modeling}, and conversational engagement \cite{sharma2020engagement}. Research has also been conducted on building virtual assistants and chatbots for behavioral health \cite{darcy2021evidence,nicol2022chatbot,sinha2022adherence} and counseling \cite{srivastava2023response}. Further, researchers have designed AI-based systems to assist mental health support providers. Tanana et al. \cite{tanana2019development} proposed a machine learning system for training counselors. Sharma et al. \cite{Sharma2021-rq,sharma2023} developed a GPT-2 and \rthree{reinforcement learning-based} system for providing empathy-focused feedback to untrained online peer supporters. Our work informs how to support self-guided mental health interventions using AI-based systems.

Prior computational work on cognitive restructuring has often relied on small-scale, wizard-of-oz style studies in controlled lab settings \cite{smith2021effective,morris2015efficacy}. Further, Sharma et al. \cite{sharma-etal-2023-cognitive} and Ziems et al. \cite{ziems2022inducing} have developed language models for automating cognitive reframing and positive reframing respectively. \rthree{Sharma et al. \cite{sharma-etal-2023-cognitive} also investigate  what constitutes a ``high-quality'' reframe and find that people prefer highly empathic or specific reframes, as opposed to reframes that are overly positive. Our work expands on these studies by investigating the design of a system that supports people through various steps of cognitive restructuring of negative thoughts. We evaluate this intervention in an ecologically valid setting with individuals experiencing mental health challenges on a large mental health platform. In addition, through randomized controlled trials, we assess the impact of key design hypotheses associated with this intervention including personalizing the intervention to the participant, facilitating iterative interactivity with the language model, and pursuing equity. Building on Sharma et al. \cite{sharma-etal-2023-cognitive}, we also develop mechanisms that allow people to actively seek specific types of reframes that are likely to be helpful (e.g., actionable) and assess its effects on different restructuring outcomes.}

In their human-centered study, Kornfeld et al.~\cite{kornfield2022meeting} sought to understand the adoption of automated \rthree{text messaging} tools. The study revealed that the participants were interested in making the tools more personalized, favored varying levels of engagement, and wanted to explore a broad range of concepts and experiences. The design of our self-guided intervention builds on these findings aiming to facilitate intervention personalization through situations and emotions of participants, iterative engagement with the language model, and to improve the equity of the intervention across different participant issues and demographics. 

Prior work has also studied mental health bias in language models. Lin et al. \cite{lin-etal-2022-gendered} investigated gendered mental health stigma present in masked language models and showed that models captured social stereotypes, such as the perception that men are less likely to seek treatment for mental illnesses. We study the disparities of our intervention's effectiveness among individuals from diverse demographics and facing different issues. We also propose a way of improving the interventions for a key subpopulation.

\subsection{Design of Human-Language Model Interaction Systems}
\label{subsec:related-human-language-model-interaction}
Broadly, our work relates to the design of human-language model interaction systems that facilitate an interactive setting in which humans can effectively engage with language models to accomplish real-world tasks \cite{amershi2019guidelines,lee2022evaluating}. Examples include systems for creative writing \cite{clark2018creative}, programming (e.g., CoPilot (\href{https://github.com/features/copilot}{github.com/features/copilot})), and brainstorming ideas (e.g., Jasper (\href{https://www.jasper.ai/}{jasper.ai})). Our work studies how such human-language model interaction can support self-guided mental health interventions. 
\section{Study Overview}
\label{sec:study-overview}

Our study was conducted over a nine-month process involving iterative ideation, prototyping, and evaluation in collaboration with mental health experts (some of whom are co-authors).

\subsection{A Case Study of Cognitive Restructuring}
\label{subsec:cognitive-restructuring}

\rthree{Cognitive Restructuring is a well-established therapeutic technique that fosters awareness of and methods for changing negative thinking patterns \cite{beck1976cognitive,beck2015cognitive}. Cognitive Restructuring has been proven to be an effective treatment strategy for psychological disorders, especially anxiety and depression \cite{clark2013cognitive}. It is a process that is central to Cognitive Behavioral Therapy \cite{beck1976cognitive}, a modality of treatment which has been demonstrated to be as effective as, or more effective than, other forms of psychological therapy or psychiatric medications \cite{hofmann2012efficacy,butler2006empirical}.}

A participant initiates this process by writing the negative thought they are struggling with. Next, they try to \textit{identify any potential thinking traps} (biased or irrational patterns of thinking) in their thought. Thinking Traps, alternatively known as \textit{cognitive distortions}, refer to biased or irrational patterns of thinking that lead individuals to perceive reality inaccurately \cite{beck1976cognitive,ding2022improving}. These typically manifest as exaggerated thoughts, such as making assumptions about what others think (``\textit{Mind reading}''), thinking in extremes (``\textit{All-or-nothing thinking}''), jumping to conclusions based on one experience (``\textit{Overgeneralizing}''), etc. Finally, participants \textit{write a reframed thought} which involves appropriately addressing their underlying thinking traps and coming up with a more balanced and helpful perspective on the situation.

As an example, consider a PhD student who has struggled with their research project and starts to worry, ``\textit{I'll never complete my PhD}''. Some possible thinking traps that this thought is falling into include \textit{Catastrophizing} (thinking of the worst-case scenario) and \textit{Fortune telling} (trying to predict the future). Addressing these thinking traps, a possible way in which the student can reframe the thought could be to say, ``\textit{I'm imagining the worst-case scenario. This project did not work out, but I was able to formulate meaningful hypotheses for my next research project.}'' This process, along with our proposed system (Section~\ref{subsec:system}), is also illustrated in Figure~\ref{fig:interface}. Here, we study how cognitive restructuring can be supported through language models.

\subsection{Study Design}
\label{subsec:study-design}

\xhdr{Platform} We used Mental Health America (MHA) for our study, a large mental health website that provides mental health resources and tools to millions of users. We deployed our system on this platform, where it was hosted with appropriate ethical considerations and participant consent process (see Privacy, Ethics, and Safety), along with other existing self-guided mental health tools.

\xhdr{Participants} Many MHA visitors are interested in mental health resources including self-guided systems. Our study participants were visitors to MHA, who chose to use our system and provided informed consent to participate in the research study. Our study comprised a total of 15,531 participants.\footnote{\rone{Overall, we had 43,347 participants in total who consented initially, but 27,816 (64.17\%) dropped out before completing the outcome survey. Analysis in the paper only include the 15,531 (35.83\%) participants who completed the outcome survey. Note that dropout rates of online self-guided tools are commonly between 70\% to 99.5\% \cite{karyotaki2015predictors,fleming2018beyond}.}} We carried out several experiments in parallel, including randomized trials with independent participants. The number of participants involved in each experiment is represented by ``N'' values throughout the paper. Participants were 13 years and older (\rthree{also, see} Privacy, Ethics, and Safety below for a discussion on our reasons for including minors).

\xhdr{Tasks and Procedures} We started our study by formulating design hypotheses for developing self-guided cognitive restructuring through human-language model interaction. This was done through feedback from users of early prototypes of the system, brainstorming with mental health experts, and leveraging pertinent findings from prior work (Section~\ref{subsec:hypothesis}). Based on these design hypotheses, we developed our final system for human-language model interaction powered cognitive restructuring (Section~\ref{subsec:system}).

We evaluated the effectiveness of our system on several psychotherapy based metrics through a field study on the MHA platform (Section~\ref{sec:effectiveness-overall}). In addition, we evaluated the importance of our design hypotheses by conducting randomized trials on the platform, explicitly ablating specific design features, such as removing psychoeducation from the system (Section~\ref{sec:impact-design-hypothesis}). Finally, we analyzed and improved the equity of our system (Section~\ref{sec:equity}).

\xhdr{Privacy, Ethics, and Safety} We designed and conducted our field study after carefully reviewing the potential benefits and risks to participants in consultation and collaboration with mental health experts. Our study, including the participation of minors, was approved by our Institutional Review Board. We included minors (those aged between 13 and 17) in this study as they represent a large, key demographic on MHA and were already frequently using similar self-guided interventions outside of our study. Therefore, we concluded that disallowing minors access to some self-guided tools would not decrease risk but potentially make an effective intervention inaccessible to a large fraction of users.\rtwo{We obtained informed consent from adults and informed assent from minors}, ensuring they were fully aware of the study's purpose, risks, and benefits (Appendix Figure~\ref{supp:fig:consent-1} and Figure~\ref{supp:fig:consent-2}). Participants were informed that they would interact with an AI-based model that automatically generates reframed thoughts, without any human supervision. Further, appropriate steps were taken to avoid harmful content generation (Section~\ref{subsec:system}), but participants were informed about the possibility that some of the generated content may be upsetting or disturbing. Also, participants were given access to a crisis hotline. Examples in this paper have been anonymized using best privacy practices \cite{matthews2017stories}. \rthree{Also, see} Section~\ref{subsec:discussion-ethics} for broader discussion on the frameworks that guided our \rthree{ethical} and safety considerations.

\rtwo{For minors, we requested and received a waiver of parental permission from the IRB. This is because discussing mental health issues with parents may pose additional risks including discomfort in disclosing psychological distress \cite{smith2022online,samargia2006foregone} and reduced autonomy \cite{wilson2012brief}. Therefore, obtaining parental permission is generally avoided in similar studies \cite{schleider2022randomized}. Also, given our online setting, obtaining parental permission is impractical and logistically difficult as we are not directly interacting with the participants. We instead obtained the assent of minors as approved by the IRB.}

\section{RQ1: How can we design a self-guided cognitive restructuring intervention that is supported through human-language model interaction?}
\label{sec:design}

\rtwo{Here, we design a novel system for human-language model interaction based self-guided cognitive restructuring.}

\subsection{Design Hypotheses}
\label{subsec:hypothesis}

The design of our system was based on hypotheses that were formulated by incorporating qualitative feedback from users of early prototypes of the system, brainstorming on design decisions with mental health experts, and leveraging relevant insights from previous mental health studies. Here, we briefly describe the design hypotheses that surfaced through this process. Our primary contributions include the design (Section~\ref{subsec:system}) and evaluation (Section~\ref{sec:effectiveness-overall}) of a system based on these hypotheses. In addition, we evaluate the impact of \Htwo, 
\Hthree, and \Hfour~through randomized trials (Section~\ref{sec:impact-design-hypothesis}). Moreover, we evaluate \Hone and \Hfive~through qualitative analysis of participant feedback (Sections~\ref{sec:effectiveness-overall} and~\ref{subsec:H5-results} respectively).

\begin{figure*}[t]
\centering
\includegraphics[width=0.9\textwidth]{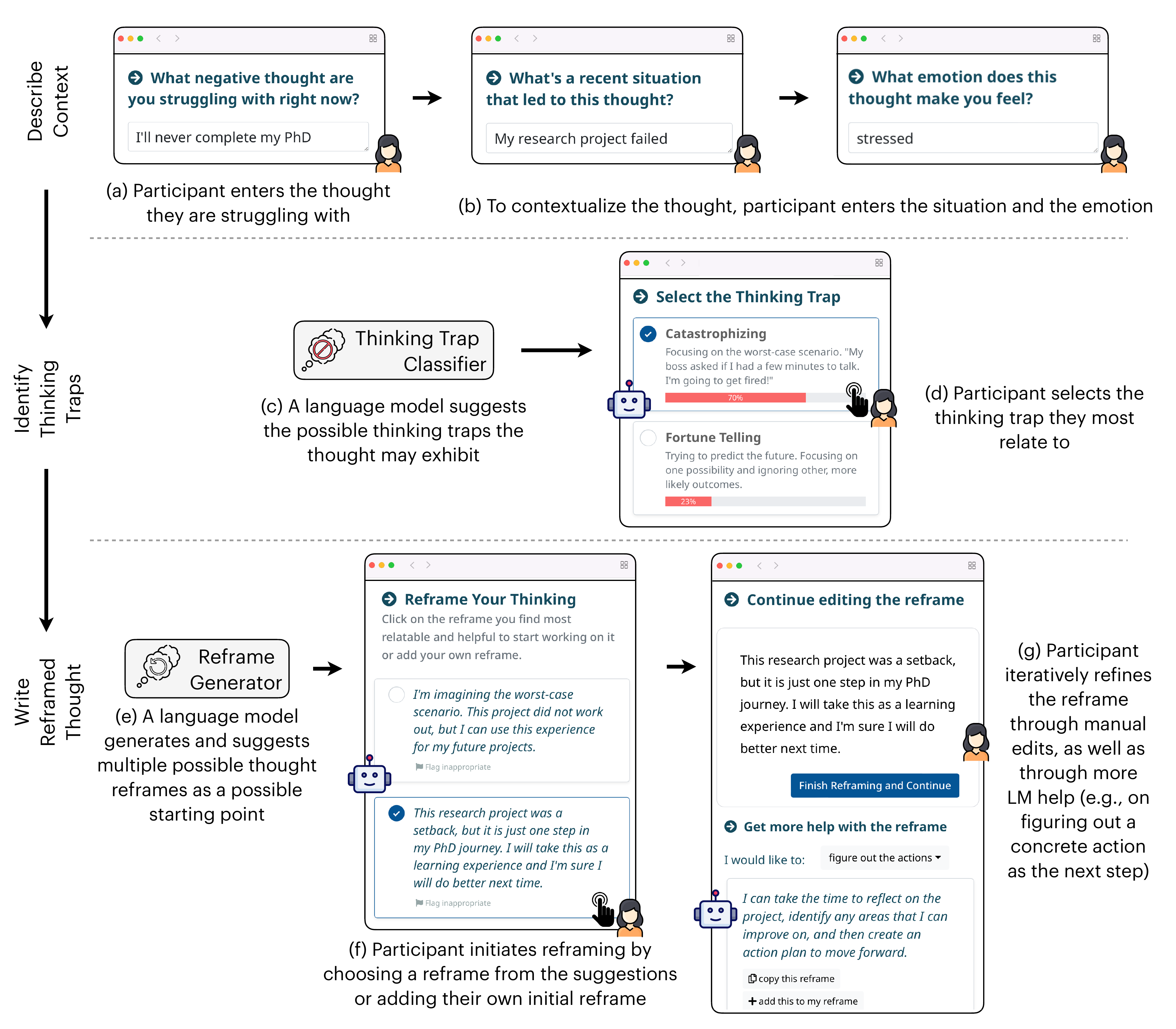}
\caption{We design a human-language model interaction based system for self-guided cognitive restructuring of negative thoughts. The system involves (a-b) describing the context by participants, (c-d) LM-assisted identification of thinking traps, and (e-g) LM-assisted writing of reframed thoughts.}
\Description{The figure illustrates the interface for our human-language model interaction based system for self-guided cognitive restructuring of negative thoughts. First row: Through the system, the participant first describes context by entering the thought
they are struggling with (e.g., "I'll never complete my PhD"). Further, to contextualize the thought, the participant enters the situation (My research project failed) and the emotion (stressed). Second row: A language model (thinking trap classifier) suggests the possible thinking traps the thought may exhibit along with their individual estimated likelihoods (e.g., “Catastrophizing – 70\%; Fortune Telling – 23\%; Overgeneralizing – 7\%”). In order to incorporate psychoeducation, we also provide definitions and examples of these thinking traps. The participant selects the thinking trap they most relate to. Third row: A language model (reframe generator) generates and suggests multiple possible thought reframes as a possible starting point (e.g., "I’m imagining the worst-case scenario. This project did not work out, but I can use this experience for my future projects." and "This research project was a setback, but it is just one step in my PhD journey.
I will take this as a learning experience and I’m sure I will do better next time."). Participant initiates reframing by choosing a reframe from the suggestions (the second one in this example) or adding their own initial reframe. Participant iteratively refines the reframe through manual edits, as well as through more LM help (e.g., on figuring out a concrete action as the next step).}
\label{fig:interface}
\end{figure*}

\xhdr{\rthree{\Hone: Assisting participants in processes that are cognitively and emotionally challenging may improve intervention effectiveness}} Previous research suggests that restructuring negative thoughts can be \textit{cognitively} and \textit{emotionally} challenging \cite{shkel2023understanding}. The limited availability of mental health professionals and resources often acts as barriers to accessibility, typically resulting in a lack of knowledge and exposure to therapeutic processes among people. Moreover, the entrenched nature of thoughts makes them difficult to overcome in the moment. In fact, many participants during our design exploration phase highlighted these challenges (e.g., a participant wrote, ``\textit{I struggle with formulating these thought --> situation --> thinking trap --> reframe scenarios by myself}''; another participant wrote, ``\textit{I find it hard to think of these reframes myself in the moment}''; another participant wrote, ``\textit{I have a hard time figuring out with cognitive distortions I'm using}''). Therefore, we aim to guide participants through the identification of thinking traps and assist them in writing effective reframes. Here, we leverage language models to achieve this, as detailed in Section~\ref{subsec:system}. Section~\ref{sec:effectiveness-overall} evaluates the effectiveness of this assistance.

\xhdr{\rthree{\Htwo: Contextualizing thought reframes through situations and emotions may improve intervention effectiveness}} Cognitive Behavioral Therapy posits that our thoughts are shaped by our situations, subsequently influencing our beliefs and emotions \cite{beck1976cognitive}. Therefore, the ability to reflect on situations and emotions and recognize their connection with negative thoughts could be beneficial in accurately assessing thinking traps and writing effective reframes. Our system enabled participants to contextualize their thoughts by answering additional questions related to their situations and emotions (Section~\ref{subsec:system}). However, it is well established that introducing additional burdens in the form of questions may lead to increased dropout, a core challenge in digital mental health \cite{baumel2019objective,torous2020dropout}. Section~\ref{subsec:H2-results} describes the randomized trial that assesses the impact of this contextualization on the tradeoff between intervention effectiveness and overall participant engagement. 

\xhdr{\rthree{\Hthree: Integrating psychoeducation may improve intervention effectiveness}} Cognitive Restructuring is a skill people can learn. Learning this skill enables participants to identify their negative thoughts, assess the thinking traps they often fall into, and develop the ability to reframe them into something more hopeful in the moment~\cite{hundt2013relationship,strunk2014assessing}. However, acquiring this skill is not straightforward and typically requires comprehensive psychoeducation coupled with practice (e.g., one participant wrote ``\textit{I don't think that everyone knows about thinking traps and the types or kinds of thinking traps there is, so I think there should be a description or definition about each thinking traps}''). Here, we introduced participants to different thinking traps through definitions and examples, along with strategies to reframe their specific thinking traps (Section~\ref{subsec:system}). Section~\ref{subsec:H3-results} details the randomized trial that evaluates the impact of integrating psychoeducation on skill learnability.

\xhdr{\rthree{\Hfour: Facilitating interactive refinement of reframes may improve intervention effectiveness}} A key component of our design involves the suggestion of automatically generated reframes through a language model (\Hone; Section~\ref{subsec:system}). However, we found that participants of early prototypes of the system desired the ability to have varied AI suggestions. One participant said, ``\textit{[I want] more choices and variation in the reframing}''. Another participant said ``\textit{I wish it included actionable insights.}'' Moreover, many participants desired the ability to interactively refine the reframes. One participant said, ``\textit{I need to write the information and revise it as many times as possible.}'' Another participant said, ``\textit{It could be more interactive, also go more in depth.}''

Therefore, our design included the option for iterative editing and updating of reframes. This enabled participants to seek more specific suggestions from the language model, including ``\textit{making it more relatable to their situation}'', ``\textit{figuring out the next steps and actions}'', and ``\textit{feeling more supported and validated}'' (Section~\ref{subsec:system}). Section~\ref{subsec:H4-results} evaluates the effects of this interactive interaction through a randomized trial.

\xhdr{\rthree{\Hfive: Mechanisms to avoid unsafe content generation and flag inappropriate content may make the intervention more safe}} The need for safety considerations is crucial when intervening in high-stakes settings like mental health \cite{li2020developing,martinez2018ethical}. There is a risk that AI might inadvertently harm, rather than help, individuals coping with mental health challenges. Therefore, a key goal was to ensure safety and minimize risks. Our design included mechanisms to avoid unsafe content generation and flag inappropriate content, as described in Section~\ref{subsec:system}. We discuss the content that was flagged by participants in Section~\ref{subsec:H5-results}.

\subsection{System Design}
\label{subsec:system}

In our system (Figure~\ref{fig:interface}), we guided participants through a five-step process. This included (1) describing the thought, (2) detailing the situation, (3) reflecting on the emotion, (4) identifying the thinking traps, and (5) finally, reframing the thought.

\xhdr{Step I, II, and III: Participant describes the context} On selecting to use the system and after consenting to participate in our study (Appendix Figure~\ref{supp:fig:consent-1} and Figure~\ref{supp:fig:consent-2}), a participant first articulates the thought they are struggling with (e.g., ``\textit{I'll never complete my PhD}''; Figure~\ref{fig:interface}a). Next, to contextualize their thought (\Htwo; Figure~\ref{fig:interface}b), the participant describes a recent situation that may have led to this thought (e.g., ``\textit{My research project failed}''). Additionally, the participant reflects on the emotion they are currently experiencing along with its intensity (e.g., ``\textit{stressed}''; 9 out of 10).

\xhdr{Step IV: LLM-assisted selection of thinking traps} The next step is to identify thinking traps. The typical process of identifying thinking traps involves participants navigating through a list to single out possible traps in their thoughts, which we identified as a cognitive and emotional barrier (\Hone). Here, we use a language model to assist participants in the identification of thinking traps among 13 common thinking traps (see Appendix Table~\ref{tab:thinking-traps}). 

For this, we rank the thinking traps for the given thoughts using a language model and show those to the participants along with their individual estimated likelihoods (e.g., ``\textit{Catastrophizing – 70\%; Fortune Telling – 23\%; Overgeneralizing – 7\%}''; Figure~\ref{fig:interface}c). In order to incorporate psychoeducation, we provide definitions and examples of these thinking traps (\Hthree; Appendix Table~\ref{tab:thinking-traps}). We use the GPT-3 model \cite{brown2020language} finetuned over a dataset of thinking traps by Sharma et al. \cite{sharma-etal-2023-cognitive}. This model achieves \rthree{a top-1 accuracy of 62.98\% on the 13-class thinking trap classification problem}. 

The participant selects one or more thinking traps from this ranked list that they most closely identify with based on their thinking pattern (e.g., ``\textit{Catastrophizing}''; Figure~\ref{fig:interface}d).

\xhdr{Step V: LLM-assisted writing of reframes} Finally, the participant writes a reframed thought addressing their thinking traps. Writing reframes to negative thoughts while maintaining composure in the moment is challenging (\Hone), therefore, we use a language model to assist participants. Our language model is based on Sharma et al. \cite{sharma-etal-2023-cognitive}, who propose a retrieval-enhanced in-context learning method using GPT-3 \cite{brown2020language}. Given a new thought $T_i$ and a situation $S_i$, this model retrieves $k$-similar examples from a dataset of \{(\textit{situation, thought, reframe}), ...\} triples collected from mental health experts. Those $k$ examples are used as an in-context prompt for the GPT-3 model to generate reframes for $T_i$ and $S_i$ ($k=5$).

Using this language model, we generate multiple suggestions for possible thought reframes and show them to the participant as a potential starting point (Figure~\ref{fig:interface}e).\footnote{To generate multiple suggestions, we perform top-p sampling \cite{holtzman2019curious} multiple times. We show three suggestions by default, but participants are provided with an option to seek more suggestions if needed.} The multiple suggestions are aimed at offering varying perspectives to the participant's original thought (e.g., ``\textit{I'm imagining the worst-case scenario. This project did not work out, but I can use this experience for my future projects.}'', ``\textit{This research project was a setback, but it is just one step in my PhD journey. I will take this as a learning experience and I'm sure I will do better next time.}'', ``\textit{I am disappointed that my research project failed, but I can still complete my PhD if I keep working hard and don't give up.}''\footnote{The reframing suggestions in this example have been generated by our GPT-3 based model.}). In addition, to incorporate psychoeducation, we provide instructions on ways in which the participant can reframe the specific thinking traps selected by them in the previous step (e.g., ``\textit{Tips to overcome catastrophizing}''; \Hthree; Appendix Table~\ref{tab:thinking-traps}).

The participant initiates reframing by choosing a reframe from the initial suggestion list or by writing a reframe on their own (Figure~\ref{fig:interface}f). The participant then iteratively refines this reframe through manual edits, as well as through additional, optional help from the language model. For this, we provide participants with an option to seek more specific suggestions from the language model, including ``\textit{making it more relatable to their situation}'', ``\textit{figuring out the next steps and actions}'', and ``\textit{feeling more supported and validated}'' (\Hfour; Figure~\ref{fig:interface}g). For the option selected by the participant, we generate additional suggestions using the language model. Participants can either copy these additional suggestions, add them to their initial reframe, replace their initial reframe with it, or use it as an inspiration. \rthree{Also, see} Appendix Figure~\ref{fig:iterative-edits} for the detailed interface.

\xhdr{Safety considerations} To minimize harmful outputs generated by language models (\Hfive), we combined classification-based content filtering with rule-based content filtering. We use a classification-based content filtering system provided by Azure OpenAI (\href{bit.ly/azure-content-filter}{bit.ly/azure-content-filter}) which identifies and filters out content related to ``hate'', ``sexual'', ``violence'', and ``self-harm'' categories. In addition, we developed a rule-based method to filter out any generated content that contained words or phrases related to suicidal ideation or self-harm. To achieve this, we created a list of 50 regular expressions (e.g., to identify phrases like ``\textit{feeling suicidal}'', ``\textit{want to die}'', and ``\textit{harm myself}'') based on suicidal risk assessment lexicons such as the one by Gaur et al., 2019 \cite{gaur2019knowledge}. A language model-generated reframe suggestion that matched any of the regular expressions was filtered out and not suggested to the participants. Also, participants were given the option to flag inappropriate reframing suggestions through a ``\textit{Flag inappropriate}'' button (Figure~\ref{fig:interface}f; Section~\ref{subsec:H5-results}).

We deployed this system on the MHA platform and studied its effectiveness with platform visitors (see Section~\ref{subsec:study-design}). We make the code used to design the system publicly available at \\ \href{https://github.com/behavioral-data/Self-Guided-Cognitive-Restructuring}{\mbox{github.com/behavioral-data/Self-Guided-Cognitive-Restructuring}}.

\section{RQ2a – To what extent does our intervention help individuals in alleviating negative emotions and overcoming negative thoughts?}
\label{sec:effectiveness-overall}

We used a mixed-method approach to evaluate the effectiveness of our system. Here, we first describe the different quantitative and qualitative measures used in our study, followed by the evaluation of our system on these metrics.

\subsection{Quantitative Measures}

Drawing from various metrics prevalent in the cognitive behavioral therapy literature, we conducted a comprehensive analysis of our system. We assessed the effects of our system on the participants' emotions, the efficacy of the reframes they wrote, and learnability of the skill:
\begin{enumerate}
    \item \textbf{Reduction in Emotion Intensity:} Intensity of the participant's emotion before the system use $-$ Intensity of the participant's emotion after the system use. We collected the emotion associated with the participant's negative thought before the system use (``\textit{What emotion does this thought make you feel?}'') and collected its intensity both before and after the system use (``\textit{How strong is your emotion? (1 to 10)}'').
    \item \textbf{Reframe Relatability:} After the system use, we asked the participant: ``\textit{How strongly do you agree or disagree with the following statement? – I \textbf{believe} in the reframe I came with}'' (1 to 5; 1: Strongly Disagree; 5: Strongly Agree).
    \item \textbf{Reframe Helpfulness:} After the system use, we asked the participant: ``\textit{How strongly do you agree or disagree with the following statement? – The reframe \textbf{helped} me deal with the thoughts I was struggling with'}' (1 to 5; 1: Strongly Disagree; 5: Strongly Agree).
    \item \textbf{Reframe Memorability:} After the system use, we asked the participant: ``\textit{How strongly do you agree or disagree with the following statement? – I will \textbf{remember} this reframe the next time I experience this thought}'' (1 to 5; 1: Strongly Disagree; 5: Strongly Agree).
    \item \textbf{Skill Learnability:} After the system use, we asked the participant: ``\textit{How strongly do you agree or disagree with the following statement? – By doing this activity, I \textbf{learned} how I can deal with future negative thoughts}'' (1 to 5; 1: Strongly Disagree; 5: Strongly Agree).
\end{enumerate}

\subsection{Qualitative Measures}
We also collected subjective feedback from participants. At the end of the system usage, we asked an optional open-ended question ``\textit{We would love to know your feedback. What did you like or dislike about the tool? What can we do to improve?}''

\subsection{Results: Quantitative}
\label{subsec:quantitative}

\begin{figure}[t]
\centering
\includegraphics[width=\columnwidth]{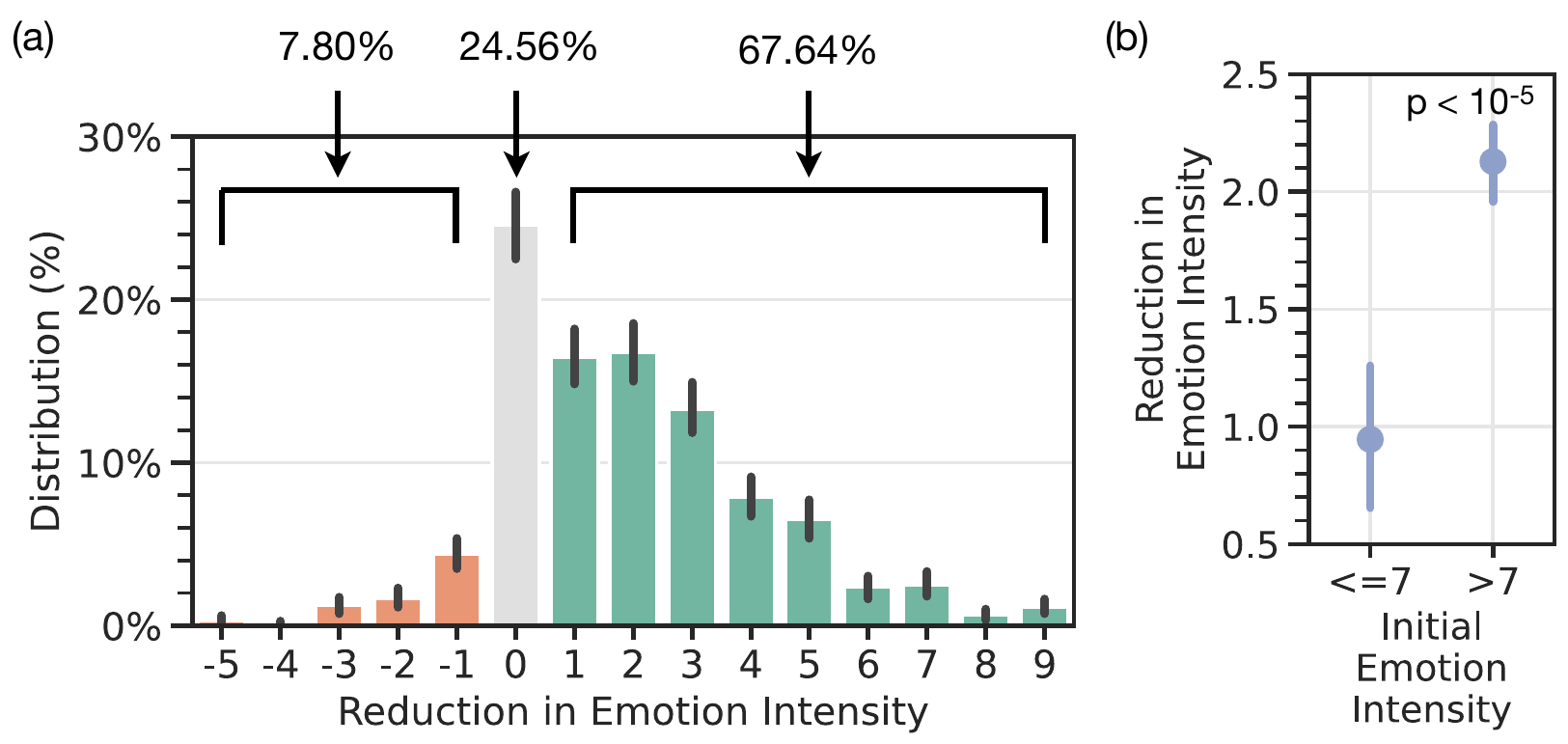}
\caption{\rthree{(a) Reduction in emotion intensity of participants before and after using the system (emotion scale: 1 to 10). We found that 67\% of the participants reported having a positive reduction in (negative) emotions (N=1,922).} (b) Participants with higher emotion intensity before using the system reported a higher reduction in emotion intensity post the system usage (N=1,922). Error bars represent 95\% bootstrapped confidence intervals.}
\Description{Two plots illustrating the effects of our system on emotion intensity. The first one is a bar plot that shows the reduction in emotion intensity of participants before and after using the system (emotion scale: 1 to 10). The x-axis is "Reduction in Emotion Intensity" and goes from -5 to 10. And the y-axis is the "Distribution" in percentage from 0\% to 30\%. 67\% of the participants report having a positive reduction in (negative) emotions, 24.56\% report no change in emotions, and 7.80\% report a negative change (N=1,922). The second one is a point plot with Initial Emotion Intensity on the x-axis (with two discrete values -- <=7 and >7) and Reduction in Emotion Intensity on the y-axis (from 0.5 to 2.5). Participants with higher emotion intensity before using the system reported a higher reduction in emotion intensity post system usage (N=1,922).}
\label{fig:emotion}
\end{figure}

\begin{table}
    \centering
    \small
    \def\arraystretch{1.15}
    \setlength{\tabcolsep}{4pt}
    {
    \begin{tabular}{lcc}

    \toprule
        \textbf{Outcome Measure} & \textbf{Mean} & \textbf{Std} \\
         \midrule
Reduction in Emotion Intensity (-10 to 10) & 1.90 & 1.29 \\ 
Reframe Relatability (1 to 5) & 3.84 & 1.17 \\
Reframe Helpfulness (1 to 5) & 3.33 & 1.35 \\
Reframe Memorability (1 to 5) & 3.52 & 1.36 \\
Skill Learnability (1 to 5) & 3.39 & 1.39 \\
 \bottomrule
    \end{tabular}}
    \caption{\rthree{Mean and standard deviation of the five quantitative measures as reported by participants.}}
    \label{tab:mean-outcomes}
    \vspace{-10pt}
\end{table}

\begin{table}
\small
\centering
\def\arraystretch{1.15}
\centering
\setlength{\tabcolsep}{4pt}
{
\begin{tabular}{lcc}
\toprule
\multirow{2}{*}{\textbf{Outcome Measures}} & \multicolumn{2}{c}{\textbf{Initial Emotion Intensity}} \\
\cmidrule(lr){2-3}
  & $<=7$ & $>7$ \\
         \midrule
Reduction in Emotion Intensity (-10 to 10) & 0.95 & 2.13 \\ 
Reframe Relatability (1 to 5) & 3.98 & 3.65 \\
Reframe Helpfulness (1 to 5) & 3.49 & 3.09 \\
Reframe Memorability (1 to 5) & 3.67 & 3.33 \\
Skill Learnability (1 to 5) & 3.57 & 3.14 \\
 \bottomrule
\end{tabular}}
\caption{\rthree{Participants with higher emotion intensity before using the system reported a higher reduction in emotion intensity post the system usage. Participants with higher initial emotion intensities reported worse reframing outcomes, suggesting that writing effective reframes and learning the cognitive restructuring skill was harder when individuals were emotionally agitated (N=1,922).}}
\label{tab:initial-emotion-intensity}
\vspace{-10pt}
\end{table}

\xhdr{67.64\% of the participants reported a positive change in emotions} We assessed the difference in the intensity of participants' self-reported negative emotions before and after utilizing our system (N=1,922). Figure~\ref{fig:emotion}a shows the distribution. Our findings revealed a positive emotional shift in 67.64\% (1,300) of the participants, while \rone{24.56\% (472) of the participants} reported no change in their emotion intensity. \rone{A small 7.80\% (150) of the participants} reported a negative shift in their emotions, with the majority (72\%; 108) of them experiencing a relatively minor negative shift of -1.

\xhdr{Participants with higher initial emotion intensity experienced a greater improvement in emotions} We checked the effects of our system on participants with different initial emotion intensities. We found that participants with more intense initial emotions (>7 out of 10) reported 124.21\% more substantial positive shifts in their emotional state than those with less intense initial emotions (2.13 vs. 0.95; <=7 out of 10; N=1,922; $p<10^{-5}$\footnote{\rone{We use a Two-sided student's t-test for all statistical tests in this paper.}}; Fig~\ref{fig:emotion}b). \rthree{Psychotherapy research suggests that a higher intensity of negative moods and depression is associated with stronger negative cognition and maladaptive thoughts \cite{beevers2007predicting}. Our findings indicate that participants with greater initial emotional intensity could have a greater benefit from a cognitive restructuring intervention like ours, potentially due to the positive effects it has on their cognition, which in turn may positively effect mood and emotion.}

\xhdr{Majority of participants found the reframes believable, helpful, and memorable} We evaluated the effectiveness of the reframes that people are able to write using our system (N=1,922). Overall, we found that 80.49\% of the participants found the reframes relatable to them, 65.65\% of participants found the reframes helpful in overcoming negative thoughts, and 70.49\% of participants found the reframes memorable or easy to remember. Further investigating people with different emotional intensities, we found that people with higher initial emotional intensities ($>7$ out of 10) reported 8.29\% lower reframe relatability (3.65 vs. 3.98; $p < 10^{-5}$), 11.46\% lower reframe helpfulness (3.09 vs. 3.49; $p < 10^{-5}$), 9.26\% lower reframe memorability (3.33 vs. 3.67; $p < 10^{-5}$), and 12.04\% lower skill learnability (3.14 vs. 3.57; $p < 10^{-5}$) than those with lower initial emotional intensities ($\leq7$ out of 10; N=1,922; \rthree{Table~\ref{tab:initial-emotion-intensity}}). This suggests that when individuals are emotionally agitated, it is harder to come up with effective reframes and learn cognitive restructuring. 

\xhdr{Most participants found the system helpful in learning cognitive restructuring} We assessed how effectively our system can be used to learn the skill of managing negative thoughts. We found that 67.38\% of the participants reported that the system helped them in learning how to deal with negative thoughts.

\rthree{We also report mean and standard deviations of the quantitative outcome measures in Table~\ref{tab:mean-outcomes}.}

\subsection{Results: Qualitative}
\label{subsec:qualitative}

On analyzing the qualitative feedback from participants, we observed that participants highlighted three key ways in which they found assistance from our system (\Hone). 

First, many participants indicated that the system helped them overcome cognitive barriers, especially when they ``feel stuck'', and doing this exercise is ``difficult'', ``on their own'' and ``in the moment.'' A participant wrote, ``\textit{My own reframes are difficult, and AI gives multiple other perspectives to consider.}'' Also, some participants reported that it helped them find ``the right words'' or ``ideas to start with.'' A participant wrote, ``\textit{Thank you for helping me to find the right words to clearly reframe a negative thought and how to apply the thought to my own thinking processes.}'' Another noted, ``\textit{I appreciated that the option of having the AI tool walk you through the reframing process step by step (e.g., by choosing the negative thought you may be experiencing + giving possible reframing ideas to start with/add more details to).}''

Second, participants expressed how the system enabled a less emotionally triggering experience. One participant wrote, ``\textit{I felt in control and more comforted that I can handle difficult situations with confidence.}'' Another participant wrote, ``\textit{This activity let me calm down...}''. Another participant noted, ``\textit{...this made the process much less daunting...}''. This is perhaps consistent with the quantitative findings on reduced emotion intensity (Section~\ref{subsec:quantitative}).

Third, participants valued that the system allowed them to explore multiple viewpoints. One participant wrote, ``\textit{...After reading several reframes and looking over them I realized that there are many options, many positive sides}.'' Another participant wrote, ``\textit{I felt reassured to see multiple views, and reflect upon them...}''

\rtwo{Overall, these results suggest that there are opportunities to assist participants in cognitively challenging and emotionally triggering psychological processes through human-language model interaction.}
\section{RQ2b – What is the impact of individual design hypotheses on the effectiveness of the intervention?}
\label{sec:impact-design-hypothesis}

Here, we studied the impact of individual design hypotheses (Section~\ref{subsec:hypothesis}) on the effectiveness of the intervention and overall participant engagement. To facilitate this, we deployed different design variations of our system by ablating specific design features (e.g., one variation that includes psychoeducation and another variation that removes it). For each design ablation, we conducted randomized controlled trials in which incoming participants were randomly assigned one of the two design variations (Appendix Table~\ref{tab:rct-options}). To measure the impact, we evaluated the difference in outcomes between participants involved in the two design variations. In this section, we report the results from ablating contextualization (\Htwo), psychoeducation (\Hthree), and interactivity (\Hfour).

\subsection{Contextualizing through Situations Improves Reframe Helpfulness}
\label{subsec:H2-results}

Reflecting on situations and emotions and understanding their connection with negative thoughts can help in writing more personalized and effective reframes (\Hone). However, asking participants for additional information like descriptions of a relevant situation and emotion can potentially increase dropout which could prevent successful outcomes. Therefore, the reflection process comes with a tradeoff with higher participant burden and higher dropout rates. Here, we conducted two different randomized trials. One where we enabled contextualization through situations to half of the participants at random. Another where we enabled contextualization through emotions to half of the participants at random.

\begin{figure}[t]
\centering
\includegraphics[width=\columnwidth]{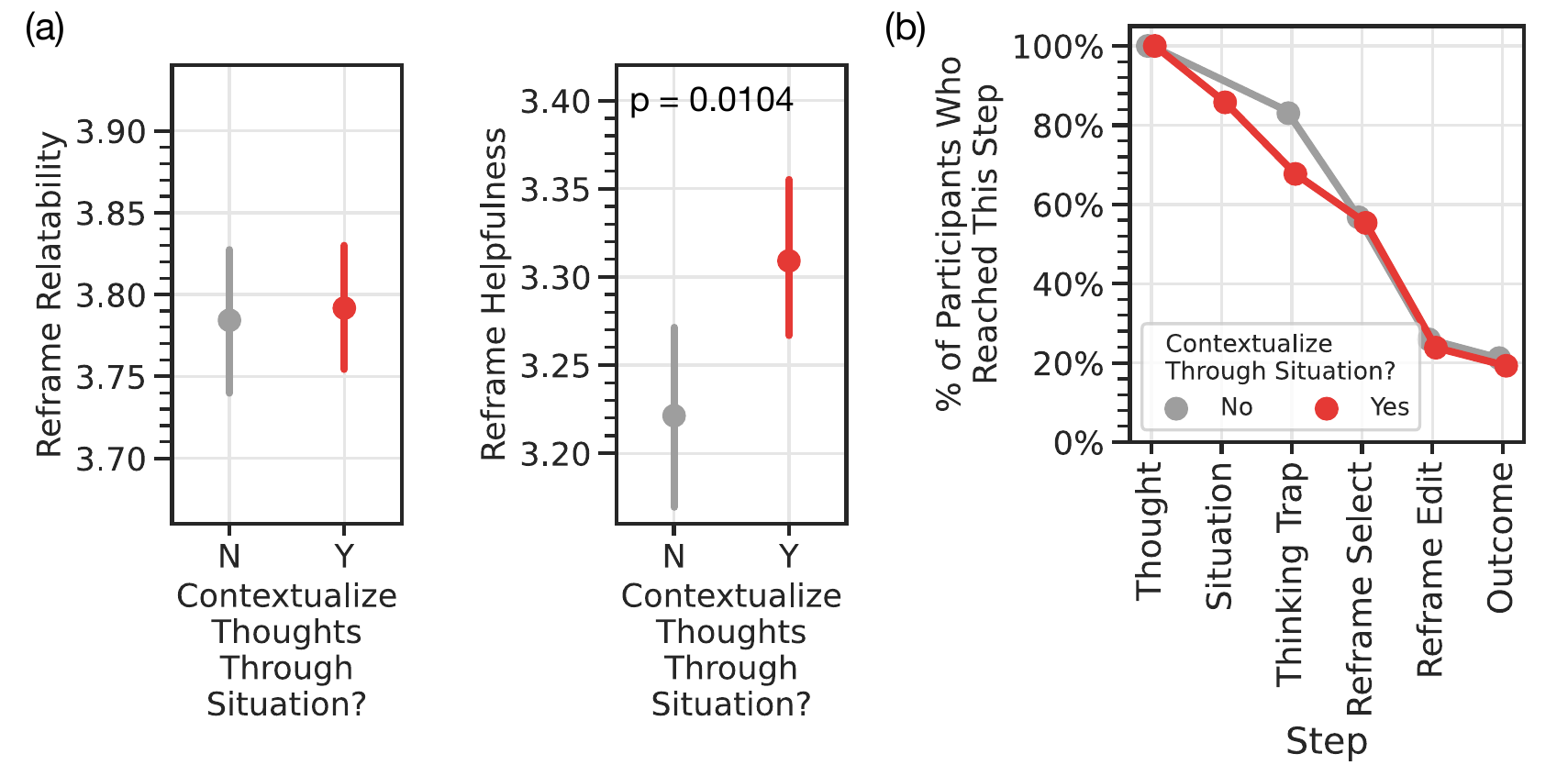}
\caption{Randomized controlled trial to estimate the effects of contextualizing thoughts through situation (N=1,636). (a) Contextualizing participant thoughts through their situations led to 2.80\% more helpful reframes (3.31 vs. 3.22; $p = 0.0104$) but did not lead to more relatable reframes. (b) Asking for additional context did not lead to a lower completion rate. Error bars represent 95\% bootstrapped confidence intervals. Effects without p-values were not significant at $\alpha = 0.05$.}
\Description{Two separate point plots and one line plot. Both the point plots have "Contextualize through Situation?" as their x-axis with two possible values "N" and "Y". The y-axes are Reframe Relatability and Reframe Helpfulness. Plots show that contextualizing participant thoughts through their situations led to 2.80\% more helpful reframes (3.31 vs. 3.22) but did not lead to more relatable reframes. The line plot represents the \% of participants who reached various steps (y-axis) against different steps of cognitive restructuring (x-axis; Thought, Situation, Thinking Trap, Reframe Select, Reframe Edit, and Outcome). There are two lines, one each for the two possible values of "Contextualize through Situation?" ("N" or "Y"). The plot shows that asking for additional context did not lead to a lower completion rate.}
\label{fig:situation-step}
\end{figure}

We found that contextualizing participant thoughts through their situations led to 2.80\% more helpful reframes and similar levels of relatability (3.31 vs. 3.22; N=1,636; \rtwo{$p = 0.0192$}; Figure~\ref{fig:situation-step}a). This indicates the benefits of increased reflection in self-guided mental health interventions. Also, while typically an increased information request is correlated with a higher dropout rate, we found that a similar number of participants reached the end of the tool, regardless of the additional information requested (Figure~\ref{fig:situation-step}b).

Surprisingly, participants who contextualized their thoughts through emotions reported 3.86\% lower levels of relatability (3.87 vs. 3.72; N=4,016; $p<0.001$; Appendix Figure~\ref{fig:emotion-vs-outcome}). This may be because our language model does not incorporate emotions while identifying thinking traps or suggesting reframes, due to the lack of relevant cognitive restructuring dataset containing self-reported emotion annotations. Consequently, this could possibly lead participants to develop unwarranted expectations where they anticipate their emotional states to be addressed in the reframing suggestions, even when they are not. Note that this is different for descriptions of situations, which the language model does take into account and typically reflects in the generated reframes. 

Qualitative feedback from participants indicated that they desired the inclusion of these steps as it helped them better process their thoughts (e.g., a participant wrote ``\textit{What made it especially helpful was being able to contextualize my feelings, which I feel allows for a more relatable reframe}'').

\subsection{Integrating Psychoeducation has Limited Impact on Overall Effectiveness}
\label{subsec:H3-results}

Providing psychoeducation with the intervention may help people learn the cognitive restructuring skill more effectively. While users of early prototypes expressed interest in integrating psychoeducation (\Hthree; Section 4.1), our randomized trial indicated that it did not lead to significant quantitative improvement in outcomes including skill learnability as self-reported by participants (at $\alpha = 0.05$; N=1,850; Appendix Figure~\ref{fig:emotion-vs-outcome}).

Nevertheless, qualitative feedback from participants indicated that they found the provided definitions, examples, and strategies helpful (e.g. one participant wrote, ``\textit{I like how the tool provided explanations}''; another participant wrote, ``\textit{I like the simple explanations and examples for each thought trap}'').

\subsection{Increased Interactivity with the Language Model is Associated with Improved Outcomes}
\label{subsec:H4-results}
\begin{figure*}[t]
\centering
\includegraphics[width=0.9\textwidth]{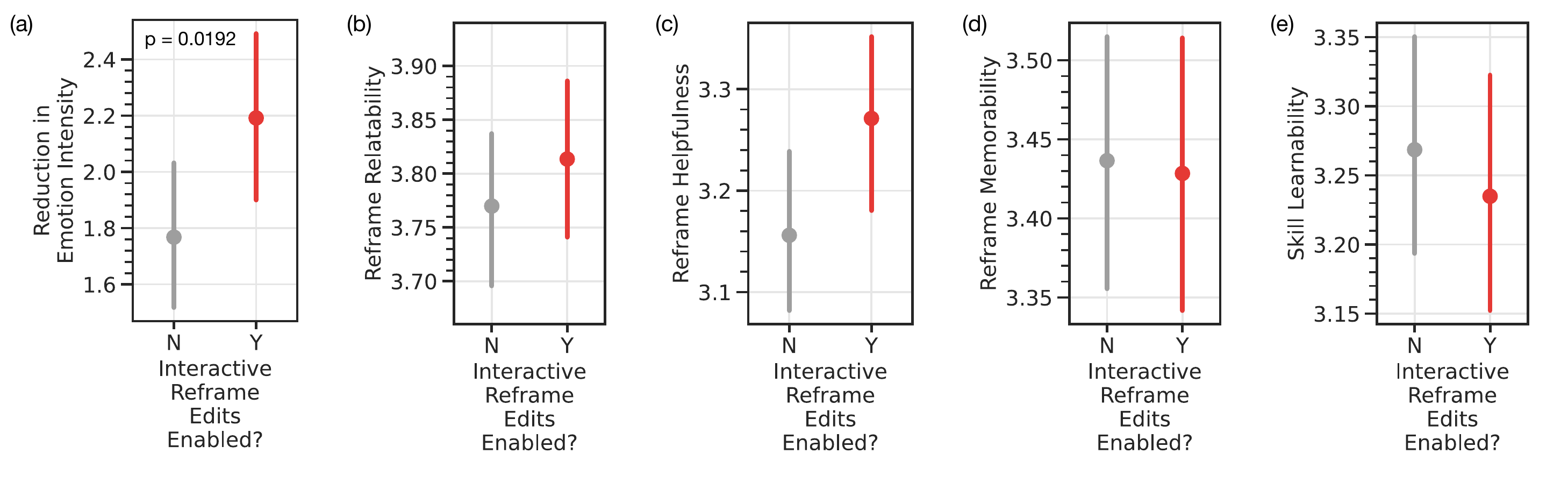}
\vspace{-10pt}
\caption{Randomized controlled trial to estimate the effects of enabling participants to iteratively edit reframes through increased interaction with the language model (N=2,165). 38\% of participants chose to use this intervention. (a) Having the option of interactive reframe edits available to participants led to a  23.73\% greater reduction in emotion intensity (2.19 vs. 1.77). (b-e) However, it did not lead to significant differences in other outcomes (at $\alpha=0.05$). Error bars represent 95\% bootstrapped confidence intervals. Effects without p-values were not significant at $\alpha = 0.05$.}
\Description{Five different points plots, all with the x-axis of "Interactive Reframe Edits Enabled?" with two possible values ("N" or "Y"). The y-axes are Reduction in Emotion Intensity, Reframe Relatability, Reframe Helpfulness, Reframe Memorability, and Skill Learnability. The plots show that having the option of interactive reframe edits available to participants led to a 23.73\% greater reduction in emotion intensity (2.19 vs. 1.77). However, it did not lead to significant differences in other outcomes (at alpha = 0.05).}
\label{fig:more-lm-rct}
\end{figure*}
\begin{figure*}[t]
\centering
\includegraphics[width=0.9\textwidth]{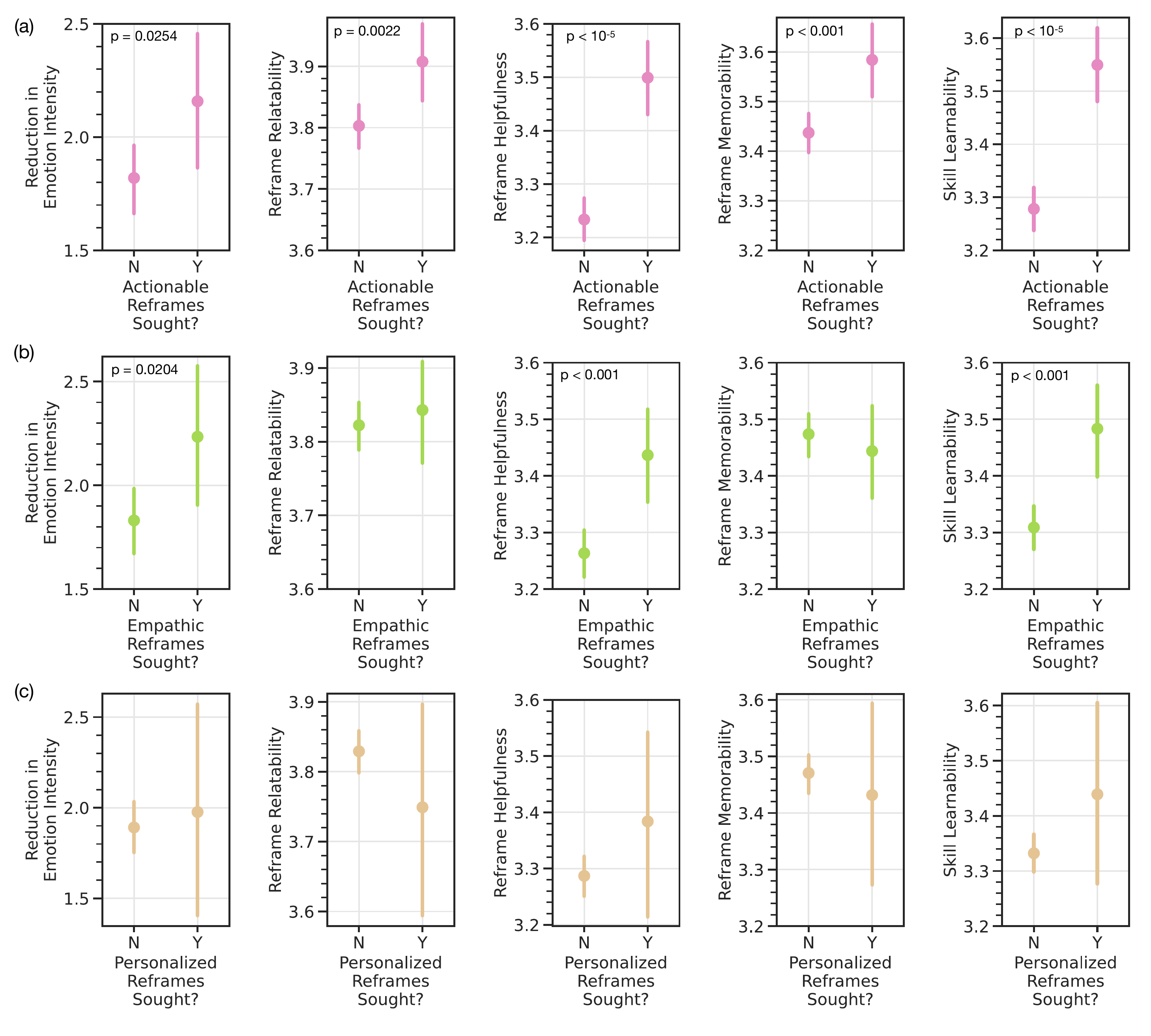}
\vspace{-10pt}
\caption{Participants were provided with an option to seek one or more of the following types of more specific suggestions from the language model -- actionable, empathic, or personalized. Among those who used any of the options (N=992), (a) those who chose to make their reframes actionable experienced superior effectiveness across all five outcomes; (b) those who chose to make their reframes empathic reported 21.86\% higher reduction in emotion intensity (2.23 vs. 1.83), 5.52\% higher reframe helpfulness (3.44 vs. 3.26), and 5.14\% higher skill learnability (3.48 vs. 3.31) and no significant differences based on reframe relatability and reframe memorability  (at $\alpha = 0.05$); (c) those who chose to make their reframes personalized reported no significant differences in outcomes (at $\alpha = 0.05$). Error bars represent 95\% bootstrapped confidence intervals. Effects without p-values were not significant at $\alpha = 0.05$.}
\Description{Three rows of five different point plots each. The x-axes for the plots in the three rows are "Actionable Reframes Sought?" ("N" or "Y"), "Empathic Reframes Sought?" ("N" or "Y"), and "Personalized Reframes Sought?" ("N" or "Y") respectively. The y-axes for the five plots in all rows are Reduction in Emotion Intensity, Reframe Relatability, Reframe Helpfulness, Reframe Memorability, and Skill Learnability. The plots show that those who chose to make their reframes actionable experienced superior effectiveness across all five outcomes. Those who chose to make their reframes empathic reported a 21.86\% higher reduction in emotion intensity (2.23 vs. 1.83), 5.52\% higher reframe helpfulness (3.44 vs. 3.26), and 5.14\% higher skill learnability (3.48 vs. 3.31) and no significant differences based on reframe relatability and reframe memorability (at alpha = 0.05). Those who chose to make their reframes personalized reported no significant differences in outcomes (at alpha = 0.05).}
\label{fig:type-of-help}
\end{figure*}

We provided participants with an option to seek more specific suggestions from the language model, including ``\textit{making the reframe more relatable to their situation}'', ``\textit{figuring out the next steps and actions}'', and ``\textit{feeling more supported and validated}'' (\Hfour; Figure~\ref{fig:interface}g; Appendix Figure~\ref{fig:iterative-edits}).

In our randomized trial, where only half of the participants at random were given this option, we found that having this option available led to a 23.73\% greater reduction in emotion intensity (2.19 vs. 1.77; \rtwo{$p = 0.0192$}; N=2,165) and insignificant differences in other outcomes (Figure~\ref{fig:more-lm-rct}). Also, some participants appreciated having this option. One participant wrote, ``\textit{I'm glad there was an option to get supported and validated}''.

Further, among the study participants who were provided this option, we observed that 38.60\% of them made use of it. Those who chose to use it to further interact with the language model to seek additional reframing suggestions of specific types (actionable, empathic, or personalized) reported 5.57\% higher reframe helpfulness (3.41 vs. 3.23; $p < 10^{-5}$) and 4.86\% higher skill learnability (3.45 vs. 3.29; $p < 0.001$) than participants who did not use it (N=992; Appendix Figure~\ref{fig:more-lm-help}). Moreover, those who chose to make their reframes actionable during this step (by choosing the option, ``\textit{I want to figure out the next steps and actions}'') reported significantly superior effectiveness across all five outcomes than those who did not (Figure~\ref{fig:type-of-help}a). Prior work has highlighted the importance of behavioral activation which involves engaging in behaviors or actions that may help in overcoming negative thoughts \cite{dimidjian2011origins,burkhardt2021behavioral}. Our work shows that participants explicitly seeking out actionable reframed thoughts are more likely to report better outcomes.

Moreover, those who chose to make their reframe empathic reported a 21.86\% higher reduction in emotion intensity (2.23 vs. 1.83; \rtwo{$p = 0.0204$}), 5.52\% higher reframe helpfulness (3.44 vs. 3.26; $p < 0.001$) and 5.14\% higher skill learnability (3.48 vs. 3.31; $p < 0.001$) and no significant differences based on reframe relatability and reframe memorability  (at $\alpha = 0.05$; N=992; Figure~\ref{fig:type-of-help}b). We did not find significant differences based on whether a participant chose to make a reframe personalized or not (at $\alpha = 0.05$; N=992; Figure~\ref{fig:type-of-help}c).

\subsection{Analysis of the Language Model Generated Content that was Flagged Inappropriate}
\label{subsec:H5-results}
We implemented a feature for participants to report any inappropriate content generated by the language model. This was achieved by including a distinct ``Flag inappropriate'' button for each generated reframe (Section~\ref{subsec:system}). Overall, we found that 0.65\% (301 out of 46,593) of the reframing suggestions shown were flagged. After conducting a qualitative review of these flagged reframes, we found a few dozen instances where the model's suggestions repeated the negative factors/sentiment described by the participant in the original thought which may inadvertently reinforce negative beliefs about oneself (e.g., ``\textit{I may be a ``failure'', but I'm still trying my best.}'' in response to the thought ``\textit{I'm a failure}''). Note that in most cases repeating parts of the participant's thought or situation helps to validate their experience and emotional reaction and personalize their reframe. Therefore, this highlights the importance of effectively differentiating which aspects of the participant's thoughts to re-state in the reframing suggestions. Future work should look more closely at how to facilitate this differentiation.

Nevertheless, for many of the 301 flagged instances, because all participants were shown three reframes as starting points, participants were able to select a different reframing suggestion from the three options presented to them, eventually reporting favorable outcomes. \rtwo{We also checked user dropout between users who flagged content vs. those who did not and found no significant differences. Interestingly, the dropout rate was slightly lower among users who flagged content, at 38.2\%, compared to 46.5\% for those who who did not.} Notably, none of the flagged instances had references to suicidal ideation or self-harm, suggesting that the safety mechanisms designed to address these concerns were likely effective (Section~\ref{subsec:system}).
\section{RQ3 – How equitable is the intervention and what strategies may improve its equity?}
\label{sec:equity}

Next, we assess how equitable our intervention is across the issues expressed by participants (Section~\ref{subsec:equity-issues}) and across participant demographics (Section~\ref{subsec:equity-demographics}). Moreover, we work towards improving equity of our system by improving its effectivness for a specific subpopulation experiencing one of the worst outcome disparities, adolescents (Section~\ref{subsec:equity-adolescents}).

\subsection{Assessing Outcomes across Issues}
\label{subsec:equity-issues}

\begin{table*}[t]
    \centering
    \small
    \def\arraystretch{1.15}
    \setlength{\tabcolsep}{4pt}
    \begin{tabular}{M{2.5cm}|cccccc}
         \toprule 
         \multirow{3}{*}{\textbf{Issues}} & \multirow{3}{*}{\parbox{1.7 cm}{\centering \textbf{Reduction in Emotion Intensity}}} & \multirow{3}{*}{\parbox{1.7 cm}{\centering \textbf{Reframe Relatability}}} & \multirow{3}{*}{\parbox{1.7 cm}{\centering \textbf{Reframe Helpfulness}}} & \multirow{3}{*}{\parbox{1.7 cm}{\centering \textbf{Reframe Memorability}}} & \multirow{3}{*}{\parbox{1.7 cm}{\centering \textbf{Skill Learnability}}} & \multirow{3}{*}{\parbox{1.7 cm}{\centering \textbf{N}}}    \\
         & & & & & & \\
          & & & & & & \\
         \midrule
Body Image & 1.42 & 3.89 & 3.20 & 3.49 & 3.38 & 71 \\ 
Dating \& Marriage & 2.05 & 3.79 & 3.20 & 3.47 & 3.33 & 328 \\
Family  & 1.99 & 3.78 & 3.26 & 3.35 & 3.34 & 170 \\
Fear  & 1.63 & \valbad{3.53} & 3.07 & 3.26 & 3.20 & 123 \\
Friendship & 1.91 & 3.65 & 3.20 & 3.48 & 3.20 & 159 \\
Habits & 1.72 & 3.98 & 3.50 & 3.52 & 3.57 & 42 \\
Health & 2.36 & 3.91 & 3.45 & \valgood{3.77} & 3.47 & 53 \\
Hopelessness & \valbad{1.11} & \valbad{3.41} & \valbad{2.66} & \valbad{3.06} & \valbad{2.84} & 70 \\
Identity & 2.54 & 4.00 & 3.55 & 3.64 & 3.09 & 11 \\
Loneliness & 1.56 & \valbad{3.43} & \valbad{2.74} & \valbad{3.03} & \valbad{2.77} & 146 \\
Money & 1.71 & 3.73 & 2.80 & 3.43 & 3.17 & 30 \\
Parenting & 2.06 & \valgood{4.19} & \valgood{3.69} & \valgood{3.97} & 3.61 & 36 \\
School & 1.94 & 3.79 & 3.20 & 3.34 & 3.13 & 181 \\
Tasks \& Achievement & 1.65 & \valbad{3.56} & 3.04 & \valbad{3.23} & \valbad{2.99} & 232 \\
Trauma & 1.42 & 3.33 & 2.58 & 3.00 & \valbad{2.50} & 12 \\
Work & \valgood{2.31} & \valgood{3.89} & \valgood{3.54} & \valgood{3.77} & \valgood{3.58} & 258 \\
\bottomrule
    \end{tabular}
    \caption{Effectiveness of our system across different issues expressed by participants. Numbers highlighted in \valingood{green} indicate outcomes that are significantly better than the population mean ($p < 0.05$). Numbers highlighted in \valinbad{red} indicate outcomes that are significantly worse than the population mean ($p < 0.05$). We found that participants who expressed \textit{Parenting} and \textit{Work} related issues reported better outcomes than the population means. Moreover, participants who expressed \textit{Hopelessness}, \textit{Loneliness}, and \textit{Tasks \& Achievement} related issues reported worse outcomes.}
    \label{tab:issues}
\end{table*}

To better determine the effectiveness of our intervention in various scenarios, we assessed the outcomes across different types of situations and thoughts that individuals might experience. We characterized participants' situations and thoughts based on the broader issues that they relate to. In collaboration with mental health experts (some of whom are co-authors), we manually labeled 500 thoughts and situations to identify the potential issues that they are associated with. The result of this iterative open-ended coding process was a set of 16 different issues expressed by participants. These include \textit{Body Image, Dating \& Marriage, Family, Fear, Friendship, Habits, Health, Hopelessness, Identity, Loneliness, Money, Parenting, School, Tasks \& Achievement, Trauma}, and \textit{Work}. See Appendix Table~\ref{tab:issues-definition} for their definitions and examples. 

We used this dataset to finetune a GPT-3 model (\texttt{text-davinci}), which achieved an accuracy of 73.00\% on a held-out set (random performance 6.25\%). We used this model to analyze the outcomes for people experiencing different issues and to identify the issues where our intervention performed better or worse (Table~\ref{tab:issues}).

We found that participants expressing \textit{Hopelessness} and \textit{Loneliness} related thoughts reported worse outcomes relative to other issues. Participants with \textit{Hopelessness} (e.g., ``\textit{I will never be better}'') reported 41.27\% lower reduction in emotion intensity (1.11 vs. 1.89; \rtwo{$p = 0.0220$}), 8.33\% lower reframe relatability (3.42 vs. 3.72; \rtwo{$p = 0.0200$}), 16.61\% lower reframe helpfulness (2.66 vs. 3.19; $p < 0.001$), 10.53\% lower reframe memorability (3.06 vs. 3.42; \rtwo{$p = 0.0169$}), and 12.07\% lower skill learnability (2.84 vs. 3.23; \rtwo{$p = 0.0119$}) than the population means. Moreover, participants with \textit{Loneliness} (e.g., ``\textit{I feel like no one is with me}'') reported 8.45\% lower reframe relatability (3.43 vs. 3.72; \rtwo{$p = 0.0031$}), 14.11\% lower reframe helpfulness (2.74 vs. 3.19; $p < 0.001$), 11.40\% lower reframe memorability (3.03 vs. 3.42; $p < 0.001$), and 14.24\% lower skill learnability (2.77 vs. 3.23; $p < 0.001$) than the population means. These differences could suggest that thoughts related to some issues are more challenging to overcome than others (as also suggested in psychology theory \cite{hawkley2010loneliness,heinrich2006clinical,beck1975hopelessness} ) or that they represent a different subpopulation. However, we also found a lower reduction in emotion intensity (i.e., an outcome measured pre- and post-intervention), suggesting that our system might have had greater difficulty in assisting these issues. In fact, some participants commented that the reframing suggestions did not work well for issues that were too complex and nuanced. One participant wrote, \textit{``It might be too simple for more complicated problems.}'' Another participant wrote, ``\textit{More complex problems need more precise results in my opinion.}'' Some participants thought that the suggestions to such complex problems were ``\textit{superficial}'', ``\textit{artificial},'' or ``\textit{hard to relate to}.'' Future iterations of the system could benefit from designing more sophisticated language modeling solutions for complex issues.

\begin{table*}[t]
    \centering
    \small
    \def\arraystretch{1.15}
    \setlength{\tabcolsep}{4pt}
    \begin{tabular}{M{2.5cm}|cccccc}
         \toprule 
         \multirow{3}{*}{\textbf{Demographics}} & \multirow{3}{*}{\parbox{1.7 cm}{\centering \textbf{Reduction in Emotion Intensity}}} & \multirow{3}{*}{\parbox{1.7 cm}{\centering \textbf{Reframe Relatability}}} & \multirow{3}{*}{\parbox{1.7 cm}{\centering \textbf{Reframe Helpfulness}}} & \multirow{3}{*}{\parbox{1.7 cm}{\centering \textbf{Reframe Memorability}}} & \multirow{3}{*}{\parbox{1.7 cm}{\centering \textbf{Skill Learnability}}} & \multirow{3}{*}{\parbox{1.7 cm}{\centering \textbf{N}}}    \\
         & & & & & & \\
          & & & & & & \\
         \midrule
 \multicolumn{7}{c}{\textbf{Age}} \\
\midrule
13--14 & 1.84 & 3.64 & \valbad{2.94} & \valbad{3.03} & \valbad{2.99} & 146 \\
15--17 & 1.64 & \valbad{3.50} & \valbad{2.68} & \valbad{3.07} & \valbad{2.83} & 149 \\
18--24 & 1.98 & 3.78 & 3.13 & 3.48 & 3.15 & 247 \\
25--34 & 2.00 & \valgood{3.89} & 3.32 & \valgood{3.70} & 3.40 & 179 \\
35--44 & 2.15 & \valgood{3.97} & 3.40 & \valgood{3.69} & \valgood{3.51} & 109 \\
45--54 & 1.64 & \valgood{3.96} & 3.32 & 3.64 & \valgood{3.56} & 71 \\
55--64 & 1.88 & 3.96 & 3.46 & \valgood{3.96} & 3.21 & 32 \\
65+ & 1.20 & 4.00 & 3.38 & 3.88 & \valgood{4.13} & 8 \\
\midrule
 \multicolumn{7}{c}{\textbf{Gender}} \\
\midrule
Female & 1.92 & 3.78 & 3.21 & 3.44 & 3.26 & 646  \\
Male & 2.19 & 3.74 & \valbad{2.95} & 3.38 & \valbad{3.04} & 258 \\
Non-Binary & 1.94 & 3.76 & 3.30 & 3.46 & 3.22 & 54 \\
\midrule
 \multicolumn{7}{c}{\textbf{Race/Ethnicity}} \\
\midrule
AIAN & 2.17 & \valbad{2.50} & 2.67 & 3.17 & 2.67 & 6 \\
Asian & 1.91 & 3.79 & 3.08 & 3.43 & 3.12 & 216 \\
Black / African Am. & 2.43 & 3.85 & 3.30 & 3.62 & 3.49 & 47 \\
Hispanic or Latino & 1.97 & 3.91 & \valgood{3.43} & 3.59 & 3.47 & 76 \\
MENA & 1.90 & 3.78 & 2.90 & \valbad{2.98} & \valbad{2.80} & 50 \\
NHPI & 2.00 & 4.40 & 4.00 & 3.60 & 3.20 & 5 \\
White & 2.05 & 3.73 & 3.12 & 3.47 & 3.22 & 438 \\
More than One & 2.83 & 3.84 & 3.29 & 3.24 & 3.16 & 38 \\
Other & \valbad{0.78} & 3.75 & 3.06 & 3.29 & 3.02 & 48 \\
\midrule
 \multicolumn{7}{c}{\textbf{Education}} \\
\midrule
Middle School & 1.80 & 3.58 & \valbad{2.89} & \valbad{2.96} & \valbad{2.79} & 120 \\
High School & 1.80 & 3.65 & \valbad{2.97} & 3.31 & 3.09 & 313 \\
Undergraduate & 2.04 & 3.79 & 3.19 & 3.54 & 3.28 & 239 \\
Graduate & 2.30 & \valgood{3.98} & \valgood{3.47} & \valgood{3.69} & \valgood{3.52} & 211 \\
Doctorate & 1.46 & 4.21 & 2.96 & \valgood{3.93} & 3.07 & 28 \\
         \bottomrule
    \end{tabular}
    \caption{Effectiveness of our system across different demographic population. Numbers highlighted in \valingood{green} indicate outcomes that are significantly better than the population mean ($p < 0.05$). Numbers highlighted in \valinbad{red} indicate outcomes that are significantly worse than the population mean ($p < 0.05$). AIAN: American Indian or Alaska Native; MENA: Middle Eastern or North African; NHPI: Native Hawaiian and Pacific Islander. We found that adolescents, males, and those with middle school education reported worse outcomes. Moreover, adults (age $\geq$ 25) and those with graduate and doctorate education reported better outcomes.}
    \label{tab:demographics}
    \vspace{-10pt}
\end{table*}


We also observed \textit{Tasks \& Achievement} related thoughts (e.g., ``\textit{I can't finish my work}'') to have 4.30\% lower reframe relatability (3.56 vs. 3.72; \rtwo{$p = 0.0299$}), 5.56\% lower reframe memorability (3.23 vs. 3.42; \rtwo{$p = 0.0274$}), and 7.42\% lower skill learnability (2.99 vs. 3.23; \rtwo{$p = 0.0071$}). Qualitative feedback from participants with such thoughts revealed that they often sought concrete actions beyond what the reframe suggestions could offer.

Moreover, those who used our system for \textit{Parenting} and \textit{Work} reported significantly better outcomes than the population means. Those with \textit{Parenting} issues reported 12.63\% higher reframe relatability (4.19 vs. 3.72; \rtwo{$p = 0.0102$}), 15.67\% higher reframe helpfulness (3.69 vs. 3.19; \rtwo{$p = 0.0145$}), and 16.08\% higher reframe memorability (3.97 vs. 3.42; \rtwo{$p = 0.0080$}). And those with \textit{Work} issues reported 22.22\% higher reduction in emotion intensity (2.31 vs. 1.89; \rtwo{$p = 0.0197$}), 4.57\% higher reframe relatability (3.89 vs. 3.72; \rtwo{$p = 0.0198$}), 10.97\% higher reframe helpfulness (3.54 vs. 3.19; $p < 0.001$), 10.23\% higher reframe memorability (3.77 vs. 3.42; $p < 0.001$), and 10.84\% higher skill learnability (3.58 vs. 3.23; $p < 0.001$).

\subsection{Assessing Outcomes across Demographics}
\label{subsec:equity-demographics}

Language modeling interventions are known to be biased toward people of specific demographics. For interventions targeting mental health, previous research has found that language models are likely to perpetuate social stereotypes, for example, under-emphasizing men's mental health \cite{lin-etal-2022-gendered}. Broadly, this corresponds to the principle of demographic parity in the fairness in machine learning literature \cite{caliskan2017semantics,mehrabi2021survey,blodgett2020language}. 

Here, we studied the difference in outcomes of our intervention across participants of different demographics. We asked participants to optionally provide demographic information, including age (ranges between 13 to 65+), gender (Female, Male, or Non-Binary), race/ethnicity (American Indian or Alaska Native, Asian, Black or African American, Hispanic or Latino, Middle Eastern or North African, Native Hawaiian and Pacific Islander, White, More than One, or Other), and education levels (Middle School, High School, Undergraduate, Graduate, or Doctorate). 

Table~\ref{tab:demographics} reports the outcome differences. We found that participants aged 17 or younger reported 4.84\%  lower reframe relatability (3.54 vs. 3.72; \rtwo{$p = 0.0091$}), 10.03\% lower reframe helpfulness (2.87 vs. 3.19; $p < 0.001$), 10.23\% lower reframe memorability (3.07 vs. 3.42; $p < 0.001$), and 10.22\% lower skill learnability (2.90 vs. 3.23; $p < 0.001$) compared to the population mean. On the other hand, those aged 25 or above reported significantly better outcomes overall. This suggests that our intervention is less effective for adolescents and more effective for adults. Section~\ref{subsec:equity-adolescents} explores improving the effectiveness of our intervention for adolescents.

Moreover, we found that male participants reported 7.52\% lower reframe helpfulness (2.95 vs. 3.19; \rtwo{$p = 0.0162$}), and 5.88\% lower skill learnability (3.04 vs. 3.23; \rtwo{$p = 0.0364$}) than the population mean. This is consistent with prior work that shows that language models are likely to be disparate toward men's mental health~\cite{lin-etal-2022-gendered}. Race or ethnicity of the participants was not consistently associated with better or worse outcomes. However, those who identified their race as ``Other'' reported a 58.73\% lower reduction in emotion intensity than the population mean (0.78 vs. 1.89; \rtwo{$p = 0.0058$}). Moreover, those who identified their race as ``Middle Eastern or North African'' reported 12.87\% lower reframe memorability (2.98 vs. 3.42; \rtwo{$p = 0.0180$}) and 13.31\% lower skill learnability (2.80 vs. 3.23; \rtwo{$p = 0.0207$}).

Finally, based on education levels, participants with a ``Middle School'' education reported 9.40\% lower reframe helpfulness (2.89 vs. 3.19; \rtwo{$p = 0.0249$}), 13.45\% lower reframe memorability (2.96 vs. 3.42; $p < 0.001$), and 13.62\% lower skill learnability (2.79 vs. 3.23; $p < 0.001$). On the other hand, those who identified as ``Graduate'' reported 6.99\% higher reframe relatability (3.98 vs. 3.72; \rtwo{$p = 0.0020$}), 8.78\% higher reframe helpfulness (3.47 vs. 3.19; $p < 0.001$), 7.89\% higher reframe memorability (3.69 vs. 3.42; \rtwo{$p = 0.0019$}), and 8.98\% higher skill learnability (3.52 vs. 3.23; \rtwo{$p = 0.0017$}). \rone{Note that age and education are strongly correlated (pearson's correlation = 0.62)}, especially for younger participants that did not yet have the time to advance their education, suggesting that the relationship between education and outcomes may be at least partially explained by age.

\subsection{Improving Intervention Equity by Improving the Experience of Adolescents}
\label{subsec:equity-adolescents}

\begin{figure*}[t]
\centering
\includegraphics[width=0.9\textwidth]{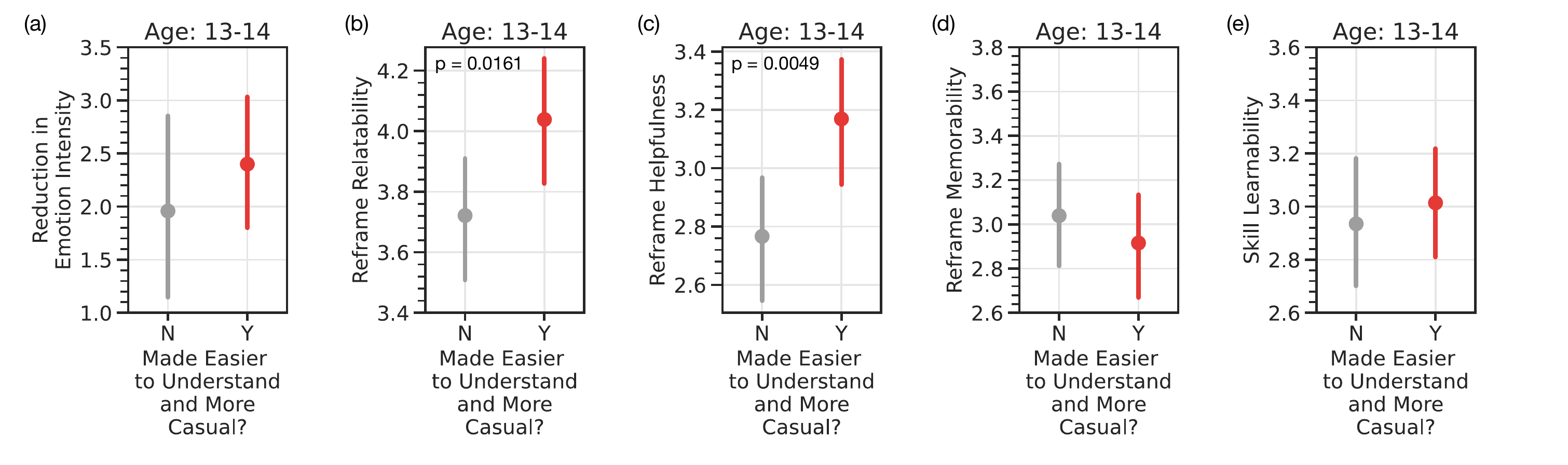}
\caption{Randomized controlled trial to estimate the effects of making reframes easier to understand and more casual on adolescents in age group 13 to 14 (N=148). Adolescents reported 8.60\% higher reframe relatability (4.04 vs. 3.72) and 14.44\% higher reframe helpfulness (3.17 vs. 2.77) if they were suggested easier to understand and more casual reframes compared to instances where such reframes were not suggested. Error bars represent 95\% bootstrapped confidence intervals. Effects without p-values were not significant at $\alpha = 0.05$.}
\Description{Five different point plots for the Age Group 13-14, all with an x-axis of "Made Easier to Understand and More Casual?" with two possible values ("N" or "Y"). The y-axes are Reduction in Emotion Intensity, Reframe Relatability, Reframe Helpfulness, Reframe Memorability, and Skill Learnability. The plots show that Adolescents in the age group 13 to 14 reported 8.60\% higher reframe relatability (4.04 vs. 3.72) and 14.44\% higher reframe helpfulness (3.17 vs. 2.77) if they were suggested easier to understand and more casual reframes compared to instances where such reframes were not suggested.}
\label{fig:age-13-14}
\end{figure*}

Because intervention effectiveness differs significantly across people's issues and demographics, it is crucial to identify solutions that improve intervention equity. This may require adapting the intervention to different subpopulations.

Here, we performed one specific experiment to study how our language model-based intervention may be adapted to make it more equitable. We particularly focused on teenagers and adolescents whom we found to have one of the largest outcome discrepancy.\footnote{While we also observed a large outcome discrepancy across educational attainment, this can largely be explained through age (as a 15 year old almost certainly did not yet have a chance to complete a college education yet).}

Research suggests that current treatment methods are often structurally incompatible with the ways adolescents engage with, or wish to engage with, mental health care \cite{kruzan2022developing}. Although adolescents are more likely to use self-guided mental health interventions \cite{schleider2020future}, our analysis suggests that our human-language modeling based intervention may be less effective for this demographic (Section~\ref{subsec:equity-demographics}). Given the escalating youth mental health crisis \cite{avenevoli2015major}, it is important to develop solutions that bridge this gap. To achieve this, we tried and identified the challenges that uniquely affect adolescents.

We hypothesized that the linguistic complexity of our system may affect its performance among adolescents. Research in sociolinguistics has shown that language use varies with age \cite{barbieri2008patterns}. On analyzing the reading complexity of the reframed thoughts authored by participants of different age groups, we found that those between the ages of 13 and 17 tend to write thoughts and reframes with the lowest levels of reading complexity (based on the Coleman–Liau Index \cite{coleman1975computer}; Appendix Figure~\ref{fig:readability-by-age}). Therefore, we tried to reduce the reading complexity of the reframing suggestions to adolescents. For this, given a reframing suggestion, we asked the GPT-3 language model \cite{brown2020language} to make it easier to understand and more casual (using the prompt, ``\texttt{\textit{Revise the following text to make it easy to understand for a 5th grader. Also, make it more casual: \{reframe\}}''}), similar to other efforts targeting people of different subpopulations (e.g., for scientific communication~\cite{august2022generating}). \rthree{Also, see} Appendix Table~\ref{tab:simpler-reframes} for examples that illustrate this rewriting process.

Figure~\ref{fig:age-13-14} reports the results of a randomized trial that only provides these easier to understand and more casual reframing suggestions to half of the participants at random. We found that adolescents in the age group 13 to 14 reported 8.60\% higher reframe relatability (4.04 vs. 3.72; \rtwo{$p = 0.0161$}) and 14.44\% higher reframe helpfulness (3.17 vs. 2.77; \rtwo{$p = 0.0049$}) when they were suggested reframes with lower reading complexity (N=148). Moreover, adolescents in the age group 15 to 17 reported 15.58\% higher reframe helpfulness (3.19 vs. 2.76; \rtwo{$p = 0.0042$}) when they were suggested such reframes (N=174). We did not find significant differences for adults ($\geq 18$) through this intervention (N=760; Appendix Figure~\ref{fig:appendix-age-15-plus}). This suggests that a simpler and more casual language might be beneficial to many. However, based on qualitative feedback, certain adult participants expressed a preference for a less casual language. Future work could explore how to accommodate such individual preferences.

\section{Discussion}
\label{sec:discussion}

\subsection{Supporting the Learning and Practice of Self-Guided Interventions}
\label{subsec:discussion-access}

Our work demonstrates how language modeling interventions can support mental health. Approximately 20\% of people worldwide are experiencing mental health problems, but less than half receive any treatment \cite{world2022mental,olfson2016building}. Due to widespread clinician shortages, lengthy waiting lists, and lack of insurance coverage, many vulnerable individuals have limited access to therapy and counseling. In addition, mental health issues are heavily stigmatized, which frequently prevents individuals from seeking appropriate care \cite{sickel2014mental}. 

Effective self-guided mental health interventions could rapidly increase access to care \cite{schleider2020future,patel2020acceptability,schleider2022randomized,shkel2023understanding}. However, despite their inherent promise, the wide-scale implementation of these interventions remains a challenge owing to the cognitive and emotional challenges that they pose \cite{shkel2023understanding,garrido2019works}. \rthree{Most interventions that digitally facilitate self-guided interventions simply transform traditional manual therapeutic worksheets into digital online formats \cite{shkel2023understanding}. These provide limited instructions and support, which affects user engagement and usage \cite{garrido2019works,baumel2019objective,fleming2018beyond,torous2020dropout}. Other studies have used wizard-of-oz methods to assist users \cite{ly2017fully,smith2021effective,morris2015efficacy,kornfield2023text,kumar2023exploring}. However, the controlled research setting of these studies limit their ecological validity, thereby limiting our understanding of user preferences when systems are deployed in real world \cite{mohr2017accelerating,blandford2018seven,borghouts2021barriers,poole2013hci}.}

\rthree{Here, we contribute the design of a novel system for human-language model interaction-based self-guided cognitive restructuring of negative thoughts. We conduct a large-scale, randomized, empirical studies in an ecologically informed setting to understand how people with lived experience of mental health interact with it.}

Our findings open up opportunities for improved learning and practicing of key mental health strategies and coping skills. Moreover, these interventions could complement traditional treatment options, e.g., by being accessible to users when they have difficulties finding a therapist, or in between sessions.

\subsection{Implications on the Design of Self-Guided Mental Health Intervention}
\label{subsec:discussion-design}

Several of our design hypotheses (Section~\ref{subsec:hypothesis}) were observed to improve intervention outcomes. These include personalizing the intervention to the participant, facilitating iterative interactivity with the language model, and pursuing equity, all of which may generalize to support other self-guided mental health interventions.

\xhdr{Personalization} Effectively supporting humans through self-guided interventions necessitates personalization \cite{kornfield2022meeting}. Our design incorporated personalization of reframes by not only seeking more information from the participants in the form of their situations, but also integrating it into the suggestions generated by the language model. We found that this form of personalization led to more helpful reframes than an intervention without it (Section~\ref{subsec:H2-results}). This was potentially beneficial in generating suggestions that were more realistic and made fewer assumptions about the participant. Moreover, it emphasized the benefits of increased self-reflection by participants, particularly when thinking about the situation associated with negative thoughts.

On the other hand, when we solicited \textit{emotions} from the participants and failed to incorporate them into the generated suggestions, we observed a significant \textit{decrease} in helpfulness (Section~\ref{subsec:H2-results}). This can likely be attributed to unrealistic expectations set up by our intervention, where participants might presume that their emotional states will be addressed in the reframing suggestions, even when that is not the case.

This shows the significance of personalizing language modeling suggestions when developing such self-guided interventions. \rtwo{Note that we did not find significant improvements based on whether participants explicitly sought specific suggestions to make a reframe personalized (Section~\ref{subsec:H4-results}). }

\xhdr{Interactivity} In our intervention, participants who interacted more actively with the language model (by seeking additional suggestions) achieved better outcomes (Section~\ref{subsec:H4-results}). This highlights the significance of designing interventions that enable a better interaction between the participant and the language model.

Moreover, qualitative feedback from participants revealed that many of them anticipated a continuous, back-and-forth interaction with the language model. Some participants desired a more in-depth exploration of thoughts. One participant wrote, ``\textit{I like it but it needs to go deeper with the thoughts.}'' Our design facilitates the iterative refinement of reframes. However, a more in-depth exploration could potentially involve addressing new thoughts that arise during the writing of a reframe for the original thought, or even tackling multiple related thoughts at once. Future iterations of the system could work towards a design that is capable of processing multiple thoughts in parallel.

Some participants even expected a chat-like interaction, probably influenced by their experiences with popular systems like OpenAI's ChatGPT (\href{https://chat.openai.com/}{chat.openai.com}). While such mechanisms may offer greater interactivity, their open-ended and uncontrollable nature creates challenges in making them conform to well-established therapeutic processes like cognitive restructuring \rtwo{\cite{stade2023artificial,choudhury2023llms,li2020developing,tate2023chatgpt}}. \rtwo{In evidence-based cognitive behavioral therapy, cognitive restructuring is typically exercised in a very structured manner (e.g., by asking a specific set of questions in a sequence). Our work shows the promise of extending such exercises with language models with similar digital interfaces.} 

\rtwo{While it is possible to replicate these steps within a pure chatbot paradigm, the nature of chat-based interactions can become more complex and may differ from the current methods employed in these interventions. For example, reviewing multiple thinking traps or integration of expandable psychoeducation content is more complex in a purely chat-based interface. Due to the similarity of our intervention with well-established therapy exercises and worksheets, our interface is likely to be a lower burden than a new chat interface. Reducing participant burden has often been associated with improved engagement and completion outcomes \cite{garrido2019works,baumel2019objective,fleming2018beyond,torous2020dropout}. In particular, this has been observed many times in our studied population (MHA platform visitors with lived experience in mental health that are not driven by study compensation incentives).} Therefore, appropriate care is required when designing interventions that offer the right trade-off between interactivity and principled adherence to theory.

\rthree{Another key aspect related to interactivity is over-reliance on the language model assistance. Taking away any ``productive struggle'' and doing the restructuring \textit{for} the user without fostering reflection and independent practice is likely unproductive. This relates to ``desirable difficulty'' that differentiates true, long-term ``learning'' of a skill from the short-term ``performance'' during skill acquisition \cite{guadagnoli2004challenge,bjork2017creating,bjork2020desirable}. A longer-term goal would be to modulate the difficulty of the self-guided intervention relative to the skill level of the user of the intervention such that the user can build their skills optimally, which forms an interesting direction of future research.}

\rthree{To assess if users are being over-reliant on a system like ours or being able to learn the skill through it, one could observe if users progressively apply the skill they are being taught in their daily lives. This would involve asking users whether they caught themselves thinking negatively, recognized the negative thinking patterns, and reframed the thought in-the-moment while they were having the thought. There exist standardized measures such as the ``Competencies of Cognitive Therapy Scale'' that operationalize this type of assessment \cite{strunk2014assessing}. While ours was a short single-use intervention, this kind of assessment requires a more longer term study, which may require different incentives, recruiting, and platforms and was therefore outside the scope of this paper.}

\xhdr{Equity} We found that our intervention was less effective for adolescents, males, and individuals with lower levels of education (Section~\ref{subsec:equity-demographics}). This is consistent with prior research, which has revealed a bias in language models used in mental health contexts toward similar demographics \cite{lin-etal-2022-gendered}. These findings highlight the importance of adapting self-guided mental health interventions utilizing AI models to suit the needs of different demographics and key subpopulations.

Our work proposes a rewriting-based method to achieve this goal (Section~\ref{subsec:equity-adolescents}). This could involve identifying the specific challenges associated with intervening in certain populations (e.g., reading complexity for adolescents) and then, designing appropriate solutions to address those challenges (e.g., lowering the reading complexity).

\subsection{Ethics and Safety}
\label{subsec:discussion-ethics}

The use of AI in mental health presents both opportunities and risks. The systematic ethical and safety considerations in our approach were based on a principle-based ethics framework, following Coghlan et al. \cite{coghlan2023chat}, Floridi \& Cowls \cite{Floridi2019Unified}, and Beauchamp \& Childress \cite{beauchamp2001principles}. 

Here, we discuss the five common principles derived from these frameworks, providing systematic guidance on how to responsibly navigate potential risks in our mental health setting.

\xhdr{(1) Non-maleficence} \textit{Avoid causing physical, social, or mental harm to participants}. We co-designed the tool with mental health experts, patient advocates, and clinicians (several of whom are co-authors of this paper) to identify any potential risks early. We further studied the safety implications prior to deployment. For this, we developed a realistic sandbox environment to ensure thorough testing instead of immediate deployment. Our safety testing involved examining system inputs that could potentially produce harmful outputs. Our safety filtering mechanisms were updated accordingly to address the potentially harmful LM-generated content (e.g., by adding specific regular expressions and using a content moderation API). Analysis of LM-generated content that was flagged suggests that these efforts were largely successful (Section~\ref{subsec:H5-results}). Additionally, throughout the study, participants were given access to a crisis hotline and were able to quit at any point. 

\xhdr{(2) Beneficence} \textit{Ensure that interventions do good or provide real benefit to participants}. This involved co-designing with mental health experts, patient advocates, and clinicians to identify opportunities to benefit. This led us to focus on cognitive restructuring, which is a well-established, evidence-based intervention that has been shown to positively impact people's mental well-being. Our large-scale field study in an ecologically valid setting demonstrates that our system positively impacts emotional intensity for 67\% of participants and helps 65\% overcome negative thoughts (Section~\ref{sec:effectiveness-overall}).

\xhdr{(3) Respect for Autonomy} \textit{Respect participants' values and choices}. We prioritized human agency and initiative by enabling participants to maintain control over what they want to write. Our language model initially offered suggestions on how to reframe thoughts. However, further interaction with the model was only possible when explicitly requested by the participant. Collection of demographic and outcome data was intentionally left optional.

\xhdr{(4) Justice} \textit{Treat participants without unfair bias, discrimination or inequity}. Our study incorporated a broad set of participants with varied demographics, including underrepresented subpopulations. We explicitly evaluated the equity of positive outcomes and identified key subpopulations where the intervention was found to be less effective. We took measures to improve the effectiveness for adolescents for whom the outcome discrepancy was among the largest observed, by lowering the reading complexity (Section~\ref{subsec:equity-adolescents}).

\xhdr{(5) Explicability} \textit{Provide users with sufficient transparency about the nature and effects of the technology, and be accountable for its design and deployment}. We sought informed consent from participants and were transparent about the use of AI, risks, and data use (Appendix Figure~\ref{supp:fig:consent-1} and Figure~\ref{supp:fig:consent-2}). We also continuously monitored quantitative and qualitative feedback for potential concerns, which also informed our design hypotheses. Finally, we make our code publicly available at \href{https://github.com/behavioral-data/Self-Guided-Cognitive-Restructuring}{github.com/behavioral-data/Self-Guided-Cognitive-Restructuring}.

\subsection{Limitations}
\label{subsec:limitations}
Our evaluation was confined to a population from a single platform. Yet, our sample size was relatively large and diverse. However, only a few participants over the age of 65 and identifying as American Indian or Alaska Native and Native Hawaiian or Pacific Islander (Table~\ref{tab:demographics}). 
While outcomes varied especially across age rather than race/ethnicity, more work is needed to identify and evaluate opportunities for culturally responsive interventions. 
The focus of this study was limited to short-term outcomes, necessitating further research to evaluate the long-term effects on participants. Still, our study contributes multiple large-scale randomized trials to inform the design and efficacy of digital mental health interventions. Our outcomes rely significantly on the quality of language model-generated thinking traps and reframes and therefore could change as language models improve. However, the algorithms and models used for generation in this paper represent the current state-of-the-art \cite{sharma-etal-2023-cognitive}.

\section{Conclusion}

In this paper, we designed a human-language model interaction system that leverages language models to support people through various steps of cognitive restructuring. Through a series of field studies and randomized trials on a large mental health website, we evaluated this system with 15,531 participants. Our findings demonstrated the effectiveness of this system in helping people reduce the intensity of their negative emotions and effectively reframing negative thoughts. Also, we proposed and validated various design hypotheses including contextualizing people's thoughts through their situations and facilitating iterative interaction with the language model. Moreover, we assessed the equity of our system across people with different issues and people of different demographics, and improved equity by demonstrably improving the experience of adolescents through lowering the reading complexity of the language model suggestions.

\begin{acks}
\label{sec:acknowledgements}
We are grateful to MHA visitors for participating in our field study. We thank the UW Behavioral Data Science Group members for their suggestions and feedback. We also thank Justin Evans for their assistance in model deployment, and Adam Miner, David Wadden, Khendra Lucas, and Sebastin Santy for their input on the tool interface. T.A., A.S. and I.W.L. were supported in part by NIH grant R01MH125179, NSF CAREER IIS-2142794, NSF grant IIS-1901386, Bill \& Melinda Gates Foundation (INV-004841), the Office of Naval Research (\#N00014-21-1-2154), a Microsoft AI for Accessibility grant, a Garvey Institute Innovation grant, and UW Azure Cloud Computing Credits.
\end{acks}

\bibliographystyle{ACM-Reference-Format}
\bibliography{_references}
\clearpage

\appendix
\section{Appendices}

\begin{table}[ht]
    \centering
    \def\arraystretch{1.15}
    \caption{Randomized Controlled Trial Experiments}
    \begin{tabular}{p{4.5cm}c}
        \toprule
        \textbf{Randomized Controlled Trial} & \textbf{Section with Results} \\
        \midrule
         Enabling contextualization through situations vs. disabling contextualization through situations & Section 6.1 \\
         \midrule
         Enabling contextualization through emotions vs. disabling contextualization through emotions & Section 6.1\\
         \midrule
         Adding psychoeducation vs. removing psychoeducation & Section 6.2 \\
         \midrule
         Enabling the option to seek more LM suggestions vs. disabling the option to seek more LM suggestions & Section 6.3 \\
         \bottomrule
    \end{tabular}
    \label{tab:rct-options}
\end{table}

\begin{figure*}[ht]
\centering
\caption{Randomized controlled trial to estimate the effects of contextualizing thoughts through emotions (N=4,016). (a) Contextualizing participant thoughts through their emotions led to 3.86\% lower relatable reframes (3.87 vs. 3.72). Note that our language model does not necessarily incorporate emotions unless they are expressed in the thought or situation as well. (b) Asking for additional context did not lead to lower completion rate. Error bars represent 95\% bootstrapped confidence intervals. Effects without p-values were not significant at $\alpha = 0.05$.}
\Description{Two different point plots and one line plot. Both the point plots have "Contextualize through Emotion?" as their x-axis with two possible values "N" and "Y". The y-axes are Reframe Relatability and Reframe Helpfulness. Plots show that contextualizing participant thoughts through their emotions led to less relatable reframes. The line plot represents the \% of participants who reached various steps (y-axis) against different steps of cognitive restructuring (x-axis; Thought, Emotion, Situation, Thinking Trap, Reframe Select, Reframe Edit, and Outcome). There are two lines, one each for the two possible values of "Contextualize through Emotion?" ("N" or "Y"). The plot shows that asking for additional context did not lead to a lower completion rate.}
\includegraphics[width=0.8\textwidth]{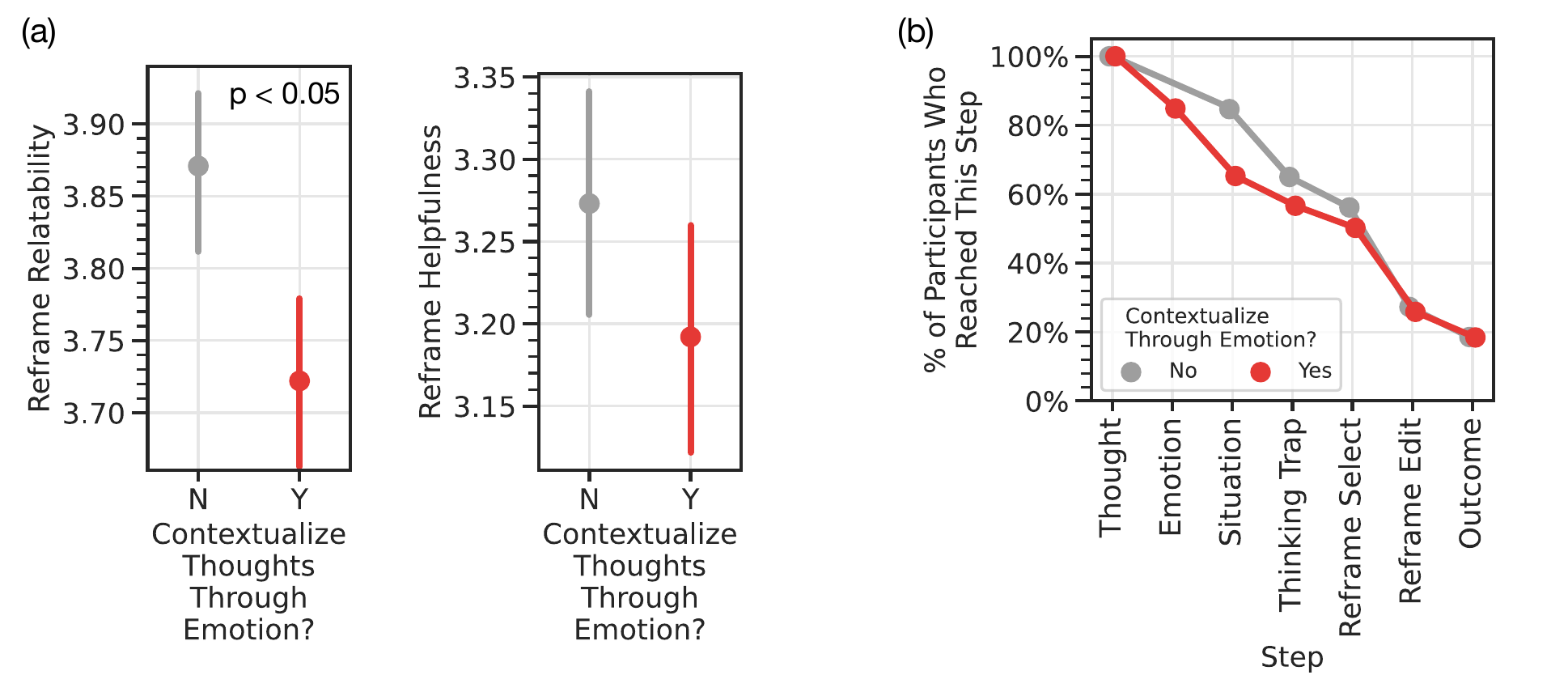}
\label{fig:emotion-vs-outcome}
\end{figure*}

\begin{figure*}[ht]
\centering
\caption{Randomized controlled trial to estimate the effects of integrating psychoeducation (N=1,850). We did not find significant quantitative improvement in outcomes on integrating psychoeducation. Error bars represent 95\% bootstrapped confidence intervals. Effects without p-values were not significant at $\alpha = 0.05$.}
\Description{Five different points plots, all with the x-axis of "Psychoeducation integrated with tool?" with two possible values ("N" or "Y"). The y-axes are Reduction in Emotion Intensity, Reframe Relatability, Reframe Helpfulness, Reframe Memorability, and Skill Learnability. The plots show that we did not find significant quantitative improvement in outcomes on integrating psychoeducation.}
\includegraphics[width=0.8\textwidth]{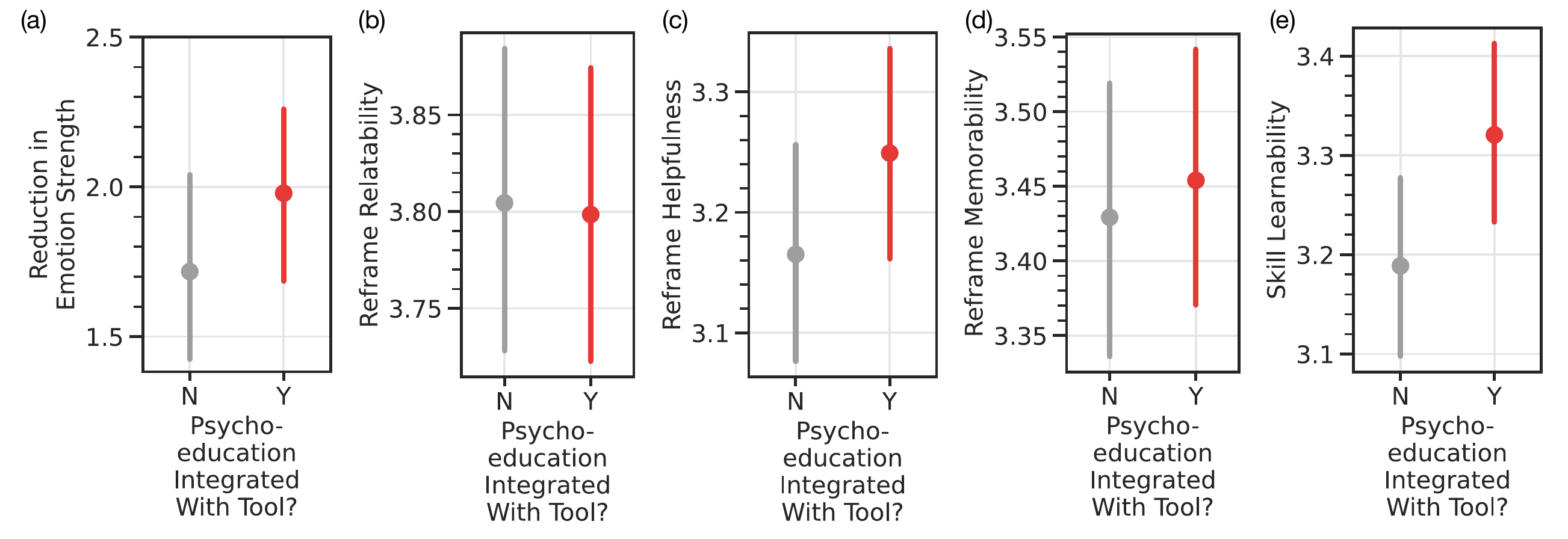}
\label{fig:psychoeducation}
\end{figure*}

\begin{figure*}[ht]
\centering
\caption{Participants who further interacted with the language model to seek additional reframing suggestions of specific types (actionable, empathic, or personalized) reported 5.57\% higher reframe helpfulness (3.41 vs. 3.23) and 4.86\% higher skill learnability (3.45 vs. 3.29) and no significant differences in reduction in emotion intensity, reframe relatability, and reframe memorability (at $\alpha = 0.05$; N=992). Error bars represent 95\% bootstrapped confidence intervals. Effects without p-values were not significant at $\alpha = 0.05$. For a randomized trial assessing the effects of this intervention, see Figure~\ref{fig:more-lm-rct}.  }
\Description{Five different point plots, all with the x-axis of "Further Interacted with LM to Seek More Help?" with two possible values ("N" or "Y"). The y-axes are Reduction in Emotion Intensity, Reframe Relatability, Reframe Helpfulness, Reframe Memorability, and Skill Learnability. The plots show that those who further interacted with the language model to seek additional reframing suggestions of specific types (actionable, empathic, or personalized) reported higher helpfulness and learnability.}
\includegraphics[width=0.85\textwidth]{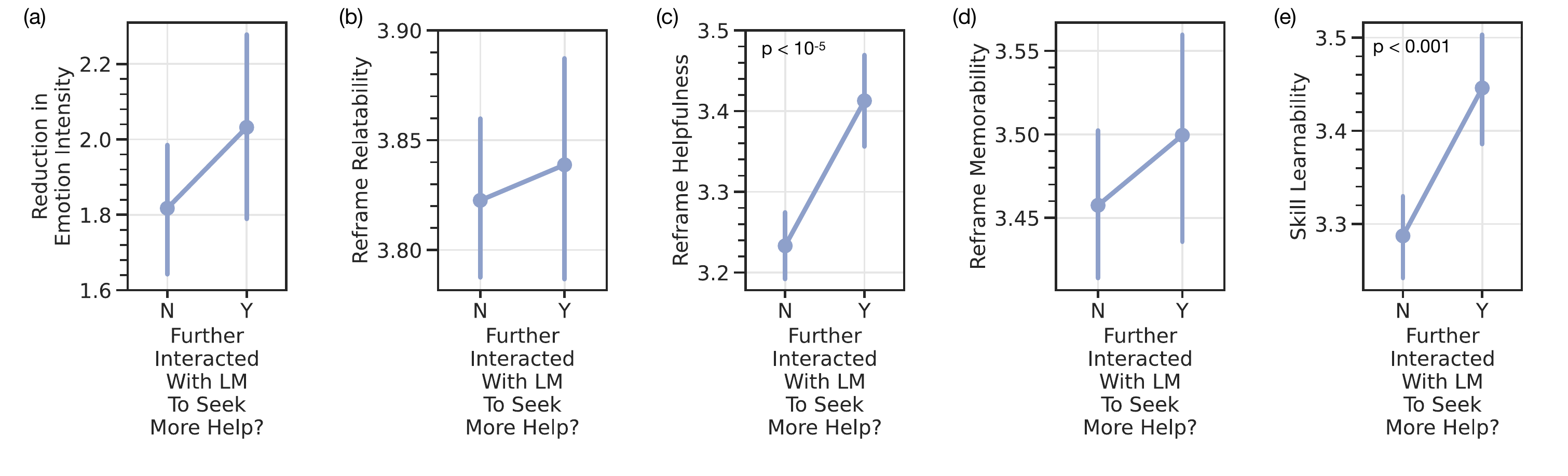}

\label{fig:more-lm-help}
\end{figure*}

\begin{figure*}[ht]
\centering
\caption{Reading Complexity (Coleman–Liau Index) of the thoughts written by participants based on their age. We find that adolescents (those below the age of 18) write thoughts with the least reading complexity. Error bars represent 95\% bootstrapped confidence intervals.}
\Description{A point plot with the x-axis as Age (with possible values of "less than 18", "18-24", "25-34", "35-44", "45-54", "55-64", "65+") and the y-axis as Reading Complexity (CLI) from 1 to 4. The plot shows that those below the age of 18 write thoughts with the least reading complexity.}
\includegraphics[width=0.5\textwidth]{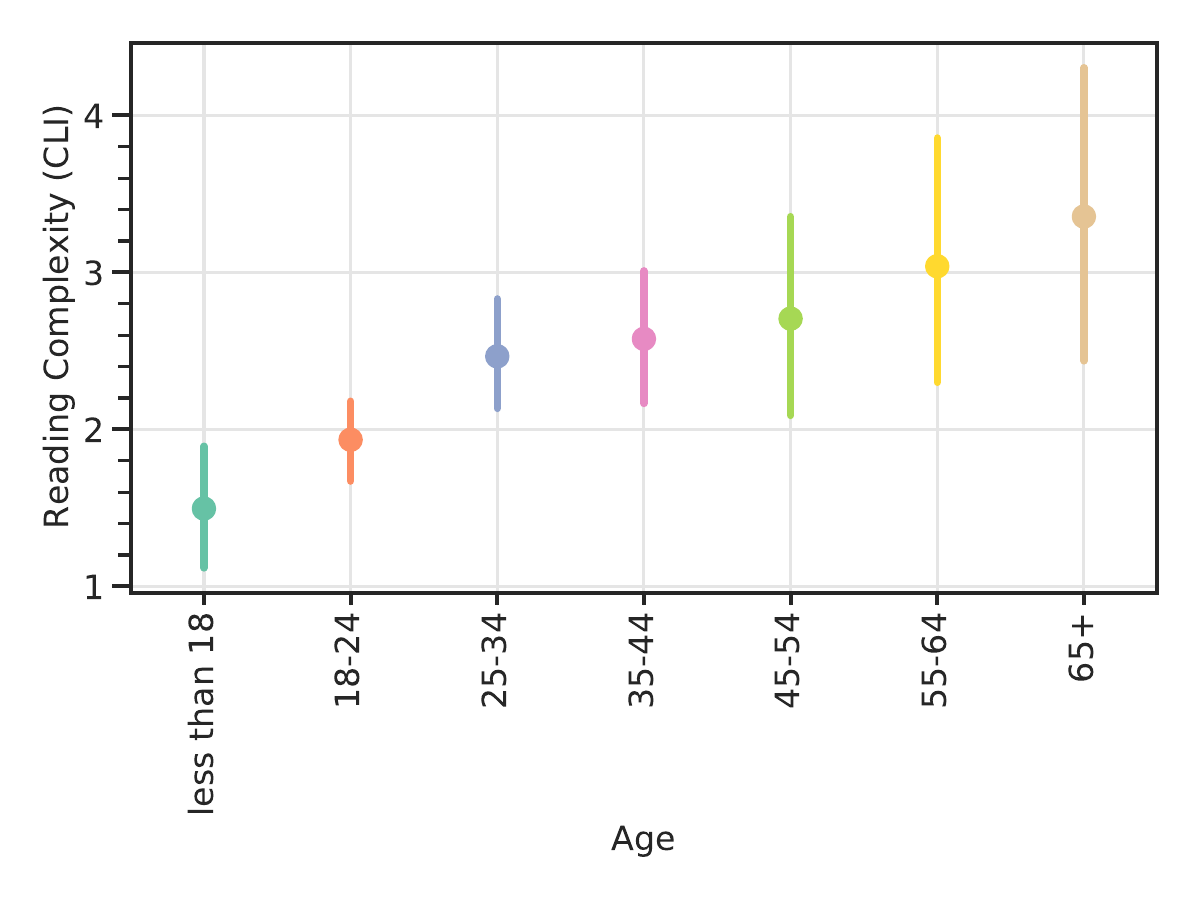}
\label{fig:readability-by-age}
\end{figure*}

\begin{figure*}[ht]
\caption{Randomized controlled trial to estimate the effects of making reframes easier to understand and more casual on age groups 15 to 17 and $\geq 18$. Participants in the age group 15 to 17 reported 15.58\% higher reframe helpfulness (3.19 vs. 2.76) when they were suggested such reframes. We did not find significant differences for participants in age group $\geq 18$ due to this intervention. Error bars represent 95\% bootstrapped confidence intervals. Effects without p-values were not significant at $\alpha = 0.05$.}
\Description{Two rows with five different point plots each. The two rows correspond to "age group 15 to 17" and "age group 18+". The x-axes for the plots are Made Easier to Understand and More Casual?" with two possible values ("N" or "Y"). The y-axes are Reduction in Emotion Intensity, Reframe Relatability, Reframe Helpfulness, Reframe Memorability, and Skill Learnability. The plots show that participants in the age group 15 to 17 reported significantly higher reframe helpfulness when they were suggested such reframes. There were no significant differences for participants in the age group 18+ due to this intervention.}
\centering
\includegraphics[width=0.95\textwidth]{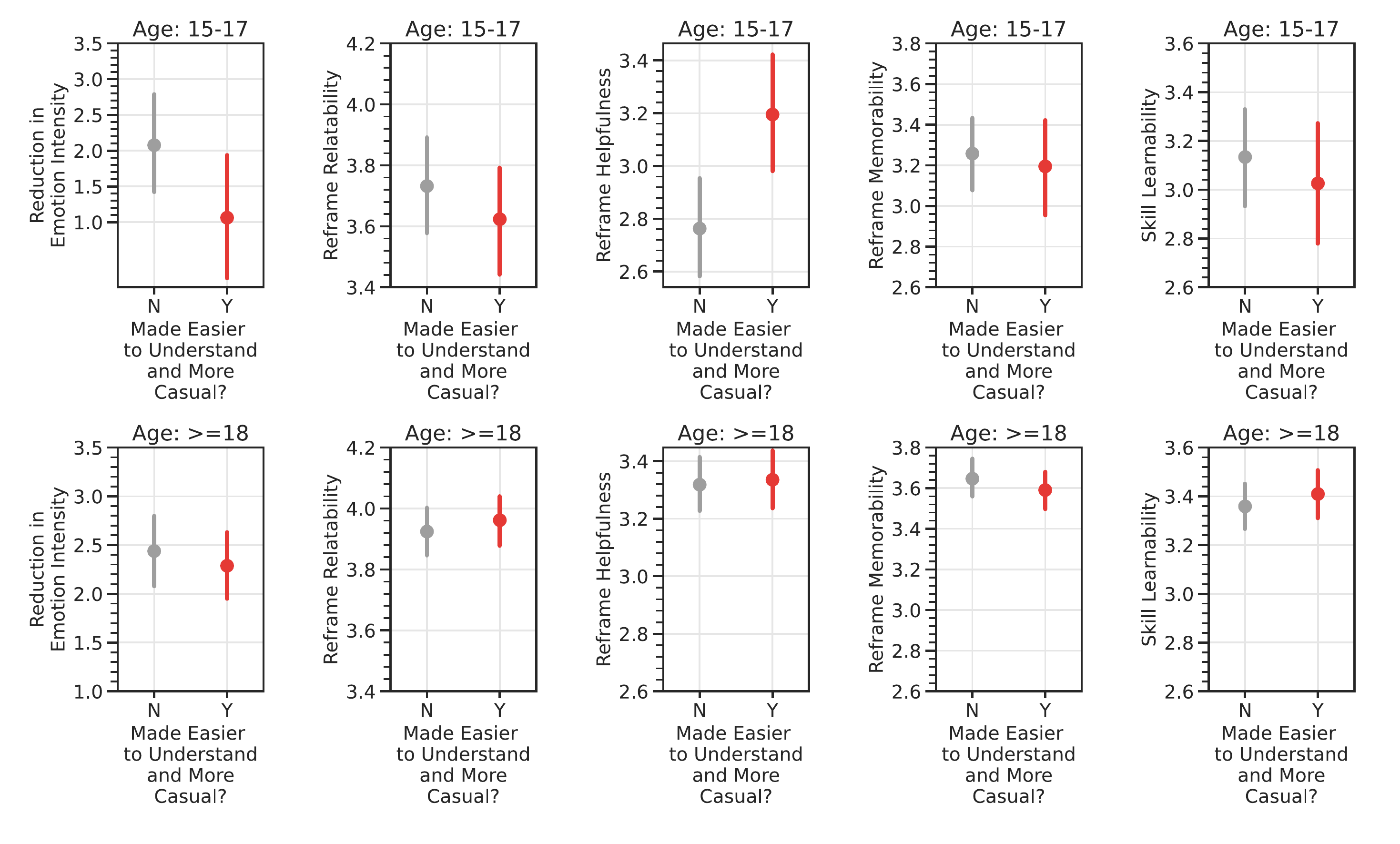}
\label{fig:appendix-age-15-plus}
\end{figure*}

\begin{figure*}[ht]
    \caption{Consent form used in our study. The form continues on the next page (1/2).}
    \Description{Page 1 (out of 2) of the consent form used in our study. The first page describes the purpose, procedure, benefits, data collection and sharing, and risks.}
    \centering
         \includegraphics[width=0.8\textwidth]{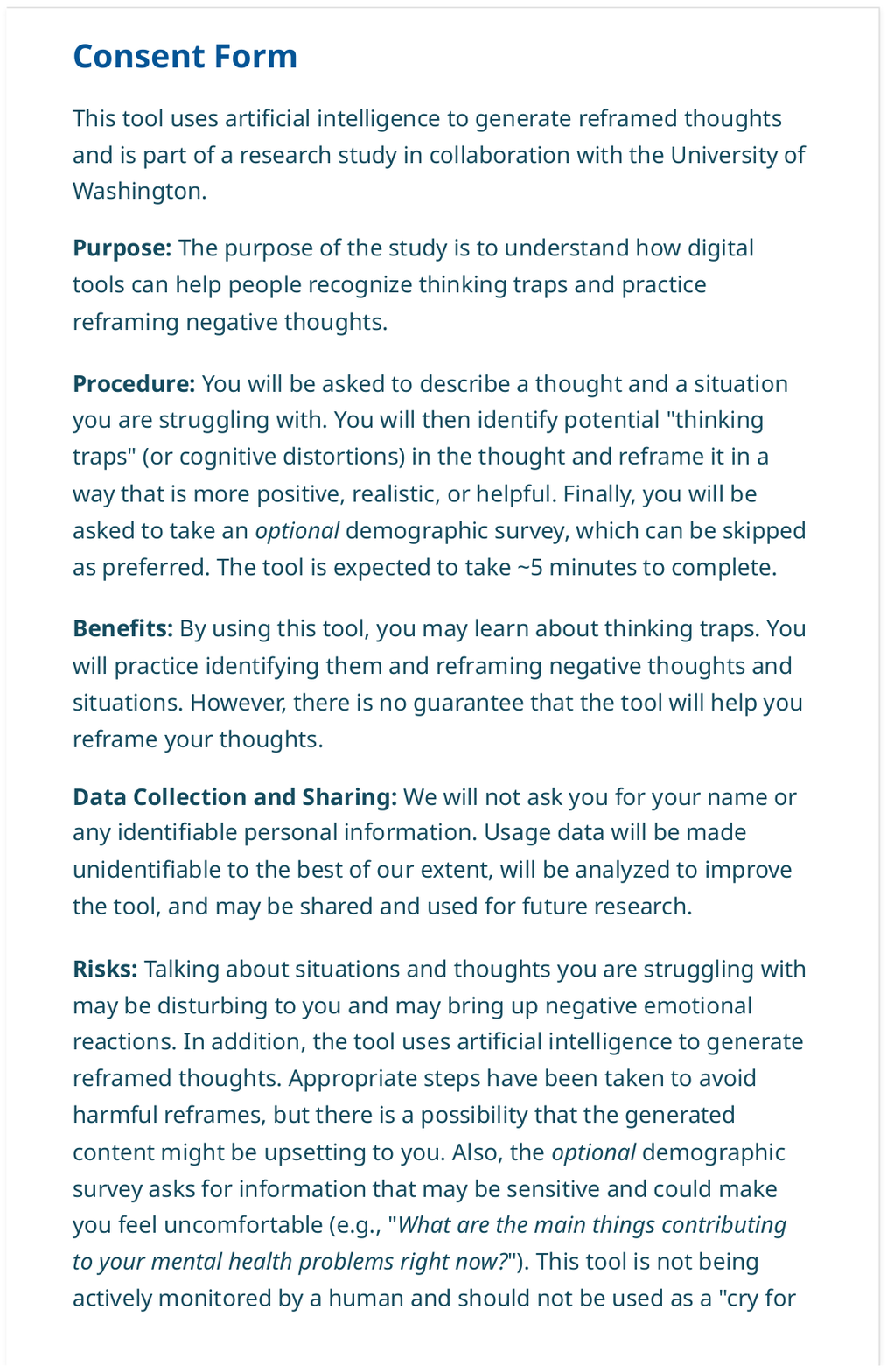}
    \label{supp:fig:consent-1}
\end{figure*}
\begin{figure*}[ht]
    \caption{Consent form used in our study (2/2).}
    \Description{Page 2 (out of 2) of the consent form used in our study. The second page describes participation details, contact information, and a checkbox for consenting.}
    \centering
         \includegraphics[width=0.8\textwidth]{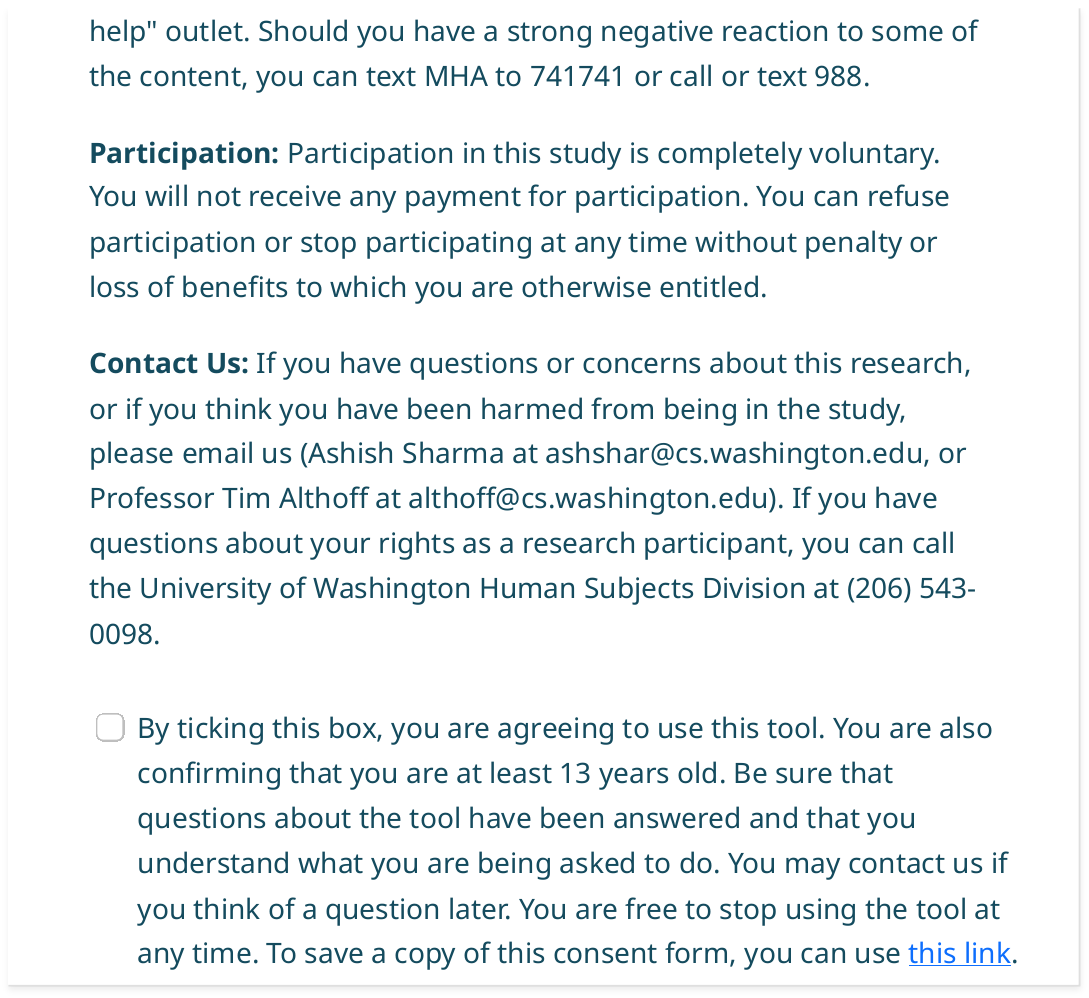}
    \label{supp:fig:consent-2}
\end{figure*}

\clearpage

\begin{table*}[ht]
    \centering
    \small
    \def\arraystretch{1.15}
\caption{Definitions, examples, and tips to overcome for each thinking trap. Definitions and examples borrowed from Sharma et al.~\cite{sharma-etal-2023-cognitive}. We also use these as part of our psychoeducation integrated into the system.}
\resizebox{\textwidth}{!}{%
\begin{tabular}{p{3cm}p{4cm}p{4cm}p{5cm}}
\toprule
\textbf{Thinking Trap} & \textbf{Definition} & \textbf{Example} & \textbf{Tips to Overcome} \\
\midrule
All-or-Nothing Thinking & Thinking in extremes. & “If it isn’t perfect, I failed. There’s no such thing as ‘good enough’.” & Things in life are rarely black and white. Focus on what’s positive or neutral about the situation. \\
Overgeneralizing & Jumping to conclusions based on one experience. & "They didn’t text me back. Nobody ever texts me back.” & Recall times when things went well for you. Imagine what it would be like for things to go well next time.\\
Labeling & Defining a person based on one action or characteristic. & "I said something embarrassing. I’m such a loser.” & Consider all different aspects of a person. \\
Fortune Telling & Trying to predict the future. Focusing on one possibility and ignoring other, more likely outcomes. & ``I'm late for the meeting. I'll make a fool of myself.'' & Be curious about what’s going to happen next. Focus on what you can control and let go of what you can’t.\\
Mind Reading & Assuming you know what someone else is thinking. & "She didn’t say hello. She must be mad at me.” &  Try to imagine other, less negative possibilities. Try to ask the person what they’re thinking, rather than just assuming.\\
Emotional Reasoning & Treating your feelings like facts. & "I woke up feeling anxious. I just know something bad is going to happen today.” & Consider all the information you have. \\
Should Statements & Setting unrealistic expectations for yourself. & "I shouldn’t need to ask for help. I should be independent.” & Think about where your unrealistic expectations came from. Let your mistakes be an opportunity to learn and grow.\\
Personalizing & Taking things personally or making them about you. & "He’s quiet today. I wonder what I did wrong.” & Think about all the other things that could be affecting someone’s behavior.\\
Disqualifying the Positive & When something good happens, you ignore it or think it doesn’t count. & "I only won because I got lucky.” & Go out of your way to notice the positive side.\\
Catastrophizing & Focusing on the worst-case scenario. & "My boss asked if I had a few minutes to talk. I’m going to get fired!” & Keep in mind that worst-case scenarios are very unlikely. Try to remind yourself of all the more likely, less severe things that could happen.\\
Comparing and Despairing & Comparing your worst to someone else’s best. & ``My niece's birthday party had twice the amount of people'' & Remember that what you see on social media and in public is everyone showing off their best.\\
Blaming & Giving away your own power to other people. &  "It’s not my fault I yelled. You made me angry!” & Take responsibility for whatever you can—no more, no less.\\
Negative Feeling or Emotion & I'm having a negative feeling or emotion which isn’t a thinking trap. & ``I am feeling lonely.'' & Feeling negative emotions is a normal part of life. Think about what we can control, and what positive things we can be grateful for.\\
\bottomrule
\end{tabular}
}
\label{tab:thinking-traps}
\end{table*}

\clearpage

\begin{table*}[ht]
    \centering
    \small
    \def\arraystretch{1.15}
\caption{Definitions and examples of the list of issues identified in our open coding process.}
\resizebox{\textwidth}{!}{%
\begin{tabular}{p{3cm}p{5cm}p{5cm}}
\toprule
\textbf{Issue} & \textbf{Definition} & \textbf{Example Thought} \\
\midrule
Body image & Feeling ugly, dieting, disordered eating & "I'm fat and ugly" \\
Dating \& marriage & Insecurities around dating and sexuality ("No one will date me" or "I got rejected") or specific situations involving a significant other ("My husband argues with me" or "My wife left me"). Does not include abuse! & "I am scared my girlfriend is going to break up with me because I dont do much for her." \\
Family & Specific situations involving interactions with family. If the only family member involved is the person's child, that's usually classified as parenting. & "I am the worst daughter in the world." \\
Fear & Intrusive thoughts about bad things happening; worrying about the future; imagining worst-case scenarios & "Something will go wrong on my flight today" \\
Friendship & Specific situations involving interactions with specific people. Doesn't have to be hyper-specific, but different from "loneliness" which is more about just the general concept of feeling isolated or unloved. & "My friend doesn't like me anymore." \\
Habits & Drugs \& alcohol, addiction, or just any habits the person is trying to break. & "I made a goal to quit smoking and failed. I keep failing at this" \\
Health & Real or imagined illness; access to healthcare. Does not include mental health. & "I might be having a serious illness" \\
Hopelessness & Feeling like things will never get better; feeling like there's no point in trying & "I have lost all hope" \\
Identity & Discrimination due to race, gender, sexual orientation, etc; or coming to terms with being LGBTQ+. & "I'm not who I want to be. I hate my appearance and my voice." \\
Loneliness & Not having friends; being isolated from loved ones; general social anxiety & "Why don't i have friends?" \\
Money & Financial troubles. Includes being jobless if the main worry is finances. & "My financial situation is going out of hand. I'm worried about my future" \\
Parenting & Feeling like a bad parent, or worried about becoming a parent & "I'm a bad mom" \\
School & Bad grades, getting into college, fears of graduating, etc. & "I will fail my exam" \\
Tasks \& achievement & Worrying about not being good enough. Examples: hobbies, chores, executive function. & "I'm not good enough" \\
Trauma & Violence, rape, verbal or physical abuse, etc. & "I can't get through this" \\
Work & Situations in the workplace. Includes being jobless if the main worry is competency. & "I'm late for the meeting. This shows what a jerk I am" \\
\bottomrule
\end{tabular}
}
\label{tab:issues-definition}
\end{table*}

\begin{table*}
\centering
    \small
    \def\arraystretch{1.2}
\caption{Examples illustrating the rewriting of reframes to make them simpler and casual for adolescents. Simpler and casual reframes use more straightforward words and phrases (e.g., ``\textit{valuable}'' --> ``\textit{awesome}'', ``\textit{things I have accomplished}'' --> ``\textit{stuff I've already done}'', ``\textit{potential I have}'' --> ``\textit{how much more I can do}'',  ``\textit{improve my physical health}'' --> ``\textit{make my body healthier}'') and use informal language (e.g., ``\textit{But you know what?}'', ``\textit{Plus...}''). This could potentially be more appealing especially to adolescents.}

\begin{tabular}{p{4cm}p{4cm}p{4cm}}
\toprule
\textbf{Thought} & \textbf{Original Reframe} & \textbf{Simpler and Casual Reframe} \\
\midrule
I'm worthless. I haven't done anything valuable in my life. & I may not have achieved all of my goals yet, but I'm still valuable and capable of doing great things. I can focus on the things I have accomplished and the potential I have to achieve even more. & I haven't reached all my goals yet, but that's okay! I'm still awesome and can do awesome things. I can think about the stuff I've already done and how much more I can do in the future. \\
\midrule
I will be alone my whole life. All of my friends are starting to be in relationships. & It's normal to feel lonely and to worry about not having a partner. I can focus on developing meaningful relationships with people, and on loving and appreciating myself. & It's totally normal to feel lonely sometimes and worry about not having a partner. But you know what? Instead of stressing about it, I can focus on building amazing friendships with people who really care about me. Plus, I can also learn to love and appreciate myself just the way I am. \\
\midrule
I'm going to fail in school and my body will never change & I may not do well in school right now, but I can still make changes to my lifestyle and work hard to improve my grades. I can also take steps to improve my physical health. & Right now, school isn't going so great for me. But that doesn't mean I can't do things to make it better. I can change the way I live and put in a lot of effort to make my grades improve. I can also do things to make my body healthier. \\
\bottomrule
\end{tabular}
\label{tab:simpler-reframes}
\end{table*}

\clearpage

\begin{figure*}[ht]
\centering
\caption{Detailed interface and process for iterative edits of reframes through further interaction with the language model.}
\Description{The figure illustrates the detailed interface of the iterative edit process. There are two subfigures. The first subfigure shows the three options participants can choose from — "make it more relatable to my situation", "figure out the next steps and actions", and "feel supported and validated". Each option enables further interaction with the language model. The second subfigure shows the additional reframing suggestions by the language model based on the option chosen. Participants have multiple ways of incorporating these additional suggestions into their original reframe by either copying the reframe, adding it to their current reframe, replacing their original reframe with it, or using it as an inspiration.}
\vspace{10pt}
\includegraphics[width=0.9\textwidth]{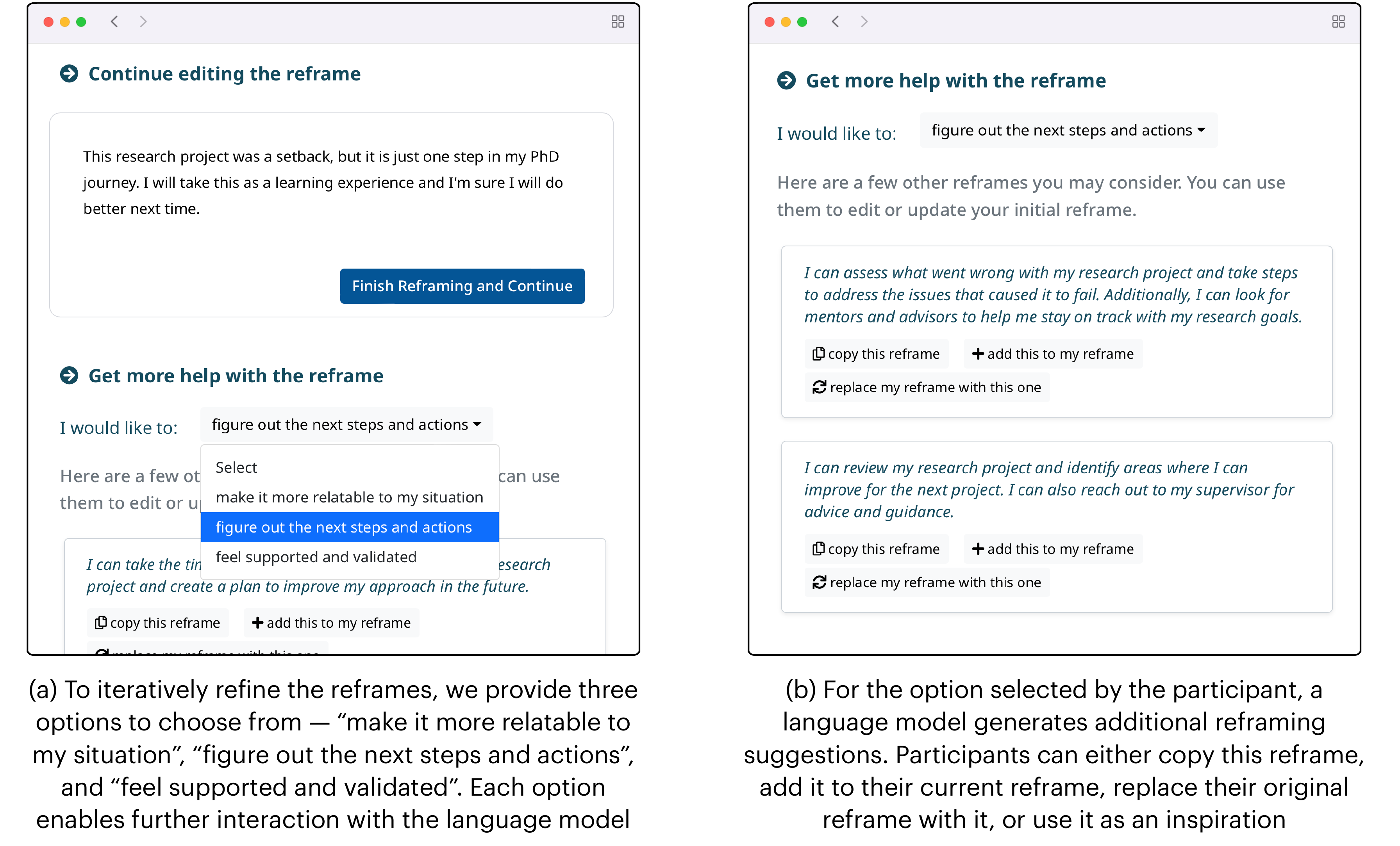}
\label{fig:iterative-edits}
\end{figure*}

\end{document}